\documentclass{elsart}
\usepackage{amsmath,amssymb,graphicx,bm}

\journal{Nonlinear Analysis}

\newcommand{\DD}[2]{\frac{\partial #1}{\partial #2}}
\newcommand{\DDD}[2]{\frac{\partial^2 #1}{\partial #2^{2}}}

\begin{document}

\begin{frontmatter}
\title{On the Nonlinear Continuum Mechanics of Space and the Notion of  Luminiferous Medium}

\vspace{-0.1in}
\author{C.I.\ Christov\thanksref{ull}\corauthref{cor}}
\ead{christov@louisiana.edu}

\address[ull]{Department of Mathematics, University of Louisiana at Lafayette, LA 70504-1010, USA}

\corauth[cor]{Corresponding author. Tel.: +1-337-482-5273; fax: +1-337-482-5346.}
      \maketitle
      
 \begin{center}
\small
Paper dedicated to the 150 anniversary of the birth of Joseph Larmor
\end{center}
     
\vspace{-0.14in}
\begin{abstract}

We prove that, when linearized, the governing equations of an incompressible elastic continuum yield Maxwell's equations as corollaries.   Through judicious distinction between the referential and local descriptions, the principle of material invariance is established and shown to be a true covariance principle, unlike the Lorentz covariance, which is valid only for non-deforming frames in rectilinear relative motion. Thus, this paper establishes that electrodynamics can be fully explained if one assumes that it is the manifestation of the internal forces of an underlying elastic material which we term the \emph{metacontinuum}.

The new frame-indifferent formulation of electrodynamics is shown to incorporate the Lorentz force as an integral part of Faraday's law, rather than as an additional empirical variable. Respectively, if the \emph{upper-convected} derivative is added in Maxwell's displacement current it can explain Biot--Savart's and Oersted--Ampere's laws.  An immediate corollary of the material invariance is the Galilean invariance of the model.

The possible detection of the absolute continuum is also discussed. First, the famous experiment of Ives and Stilwell is reexamined with a modified Bohr-Rydberg formula for the emitted frequencies from a moving atom, and it is shown that the results are fully compatible with the presence of an absolute medium. Second, a new interferometry experiment is proposed in which the first-order Doppler effect can be measured, and thus the presence of a medium at rest can be unequivocally established. 
\end{abstract}

\end{frontmatter}

\section{Introduction. Is there an Absolute Continuum?}

Any dynamic wave phenomenon is associated with either the transverse or longitudinal elastic vibrations of a medium/field; if there is a wave, something material should be `waving.' This notion led 19th century scientists to introduce the concept of the luminiferous continuum (field, aether, etc.\cite{JCM3}) (see, the illuminating account of this line of thought in \cite{Larmor_book}) .  The first attempt to explain the propagation of light as a field phenomenon was by Cauchy around 1827 (see the account in \cite{Whittaker}) who postulated the existence of an elastic continuum, through which light propagates as an elastic shear wave. It is not widely known that on the occasion of creating his model for the aether, Cauchy introduced the concept of stress tensor, which is a fundamental concept in mechanics of the continuum. Unfortunately, Cauchy's model of an elastic aether contradicted the natural perception of a particle moving \emph{through} the medium. As a result, his concept did not receive much attention, possibly because the notion of elastic liquids was not available at that time. The fluid model for the luminiferous medium was initiated by Lord Kelvin in the first half of the Nineteen Century (see the in-depth review in \cite{Wise}) and still attracts attention (see, e.g., \cite{Stevenson}). The main objection to a model based on an ideal compressible liquid is that light is a shear phenomenon, while the fluid waves are compressional in nature.

Subsequently came the contributions of Oersted, Ampere, and Faraday, eventually leading to the formulation of the electromagnetic model. The crucial advance was achieved, however, when Maxwell \cite{JCM1} added the term $\frac{\partial \bm{E}}{\partial  t}$  in Ampere's law,  which he termed the `displacement current'. This new term was similar to the time-derivative term in Maxwell's constitutive relation for elastic gases \cite{JCM2} (see also \cite{JordanPuri2}  for an insightful discussion on viscoelastic models).  Since the electric field vector is an analog of the stress vector in continuum mechanics (see \cite{Chri_FP}), one can say that Maxwell postulated an elastic constitutive relation by adding the displacement current to Ampere's law. Indeed, the new term added by Maxwell transformed the system of equations, already established in electrostatics, into a hyperbolic system with a characteristic speed of wave  propagation.  Maxwell identified the characteristic speed of this hyperbolic system with the speed of light, and thus paved the way to understanding the electromagnetic wave phenomena.

The first attempt to detect the absolute medium were based on interferometry (the famous experiment of Michelson and Morley \cite{Michelson1,MiMor})  but it produced a nil result. Many were quick to interpret the nil result of the Michelson and Morley experiment as evidence that the absolute continuum did not exist. Actually, the only thing which the nil result proves is that there is a contraction of the lengths in the direction of motion as suggested by Lorentz and FitzGerald. Later on, Lorentz found that the contraction is also a corollary from what is now called the `Lorentz transformation' (LT). Thus LT, as a mathematical tool,  produced a breakthrough result: it pointed to a possible contraction of moving bodies.

One of the most unfortunate developments in twentieth century physics came from the fact that the LT hinted at some superficial symmetry between two inertially moving frames. From the point of view of LT \cite{Ein1905}, the inertial frames appeared somehow equivalent and/or indistinguishable.  This prompted Poincare to generalize the idea of Galileo and to state the `Relativity principle' which postulates that the inertial motion of a frame cannot be detected by measurements in the same frame (see the account in \cite{Pathria}). While for a system of discrete  material points in an empty geometric space, such an assertion does not raise suspicion  (Galileo's RP), in the case of material continuous frames it is quite a stretch,  because what is obviously true for points and/or systems of points in an empty geometrical space is not necessarily true for the interacting material points of the material space (field) which points interact through the internal stresses. In fact, because light \emph{propagates} in the  absolute  continuum, while a frame is moving with respect to the latter, one should be able to measure the Doppler effect of electromagnetic waves that are emitted from a source that is at rest with respect to the absolute continuum. This is exactly what was reported in the experiments on the so called Local Standard of Rest \cite{CoreyWilk,SmootGoreMiller} (LSR).\footnote{ Apparently, coining the new euphemism `Local Standard of Rest' for the absolute continuum is a way to avoid confrontation with the current orthodoxy in physics.}

The Relativity Principle (RP) was aggressively promoted by Einstein despite the fact that Lorentz himself never embraced Poincare-Einstein brand of relativity. The infatuation with relativity principle (and more specifically with the fascinating symmetry of LT), caused scientists to act as if Maxwell's equations were untouchable. Quite ironically, ten years prior to the advent of RP, the sacred equations had already been changed  in the correct direction by Hertz \cite{Hertz}, who proposed to use the convective derivatives in the terms where Maxwell had merely partial time derivative (see, also the survey in \cite{Pinheiro}). Instead of making the smallest step to accept that Hertz's equations were valid also \emph{in vacuo},  Einstein and Minkowski took the radical approach to introduce the space-time as the model that somehow ensures the invariance.

Yet, upon closer scrutiny, the postulate of relativity does not automatically appear to be justified by the success of the Lorentz transformation. As elucidated by  Brillouin \cite{Brillouin}, relativity theory is fraught with internal logical inconsistencies and riddled by paradoxes.  Perhaps, the most obvious and mind-boggling inconsistency is the so-called twin paradox in which one observer is at rest, while the other travels in space with a given speed. Because of the indistinguishability of the inertial frames according to RP, both twins are supposed to experience the \emph{same} time dilation. It is quite reasonable to assume that the oscillatory processes (such as atomic clocks) can slow down (change their frequencies) in a frame that is moving relative to the absolute medium where the frequencies are excited. The logical absurdity, however, is to claim that each of the local times is supposed to be slowed relative to each other!  And this is exactly what follows from a rigorous application of the RP. Then the question arises: which time is actually slowed?  In other words, the question is about which of the twins will be older upon the return of the traveler. This is a \emph{purely kinematic} paradox rooted in the perplexed nature of local time embodied in the LT; however, many people accepted the  most unsatisfactory explanation based on acceleration (dynamics) which was consequently provided by Einstein. His argument was that the time slows in the frame that experiences acceleration. Somehow, the eclectic explanation was not challenged by the majority of scientists and this, most crippling, paradox  gradually became `old news' and faded away in the collective memory of science, without receiving a consistent explanation. There are many other inconsistencies when Minkowski space-time is used. For a illuminating discussion of the paradoxes of the Special Theory of relativity, see  \cite{Brillouin}.

All this means that while LT and Minkowski space are legitimate mathematical constructs, the relativity principle is still not justified. Actually, the recent measurements of LSR (see, e.g. \cite{CoreyWilk,SmootGoreMiller})  show that there must exist an asymmetry between the moving frames, depending on their speed with respect to the absolute space (preferred frame). Even if the motion is inertial, it should be detectable inside the frame because it \emph{does make a difference} who is moving relative to the underlying absolute continuum. The detection of LSR means that one can safely assume that the motion of an \emph{inertial} frame \emph{can} be detected inside the frame, if there is already an established pattern of propagating waves in the underlying continuum.  This is due the fact that space is a material continuum, rather than an empty geometric vessel. The results on LSR directly disprove Poincare's  RP.

In the present work we search for the proper model of the absolute continuum, because the  `ethereal,' gas-like substance considered in the Nineteenth century  cannot explain the wave phenomena in electrodynamics since light is a shear wave while in inviscid (and inelastic in the sense of shear elasticity) gasses, the waves are compressional (longitudinal). We discuss here the   material invariance (\emph{frame indifference}) of the absolute continuum (material space) as the true covariance in physics to replace the concept of  Lorentz covariance, because the latter is valid only in non-deforming frames in rectilinear, non-accelerating translation. Lorentz covariance is, in fact, a kind of palliative substitute for the real material invariance.  The material invariant formulation given here has the same form in any accelerating and even \emph{deforming} material frame.

\section{Lorentz Transformation: Dead-End in the Quest for Invariance}\label{sec:dead_end}

Invariance is the Holy Grail of modern physics. Since Galileo, the question of invariance of any physical description with respect to changing the inertial frame of reference (coordinate system), became the litmus test for the internal consistency of the model. Soon after Maxwell formulated his equations, it was discovered that his model was not invariant with respect to translational motion of the coordinate system. Voigt (see \cite{ErnstHsu,Macrossan}), and independently Larmor \cite[pg.228-229]{Larmor} and Lorentz \cite{Lorentz95}, spotted the fact that the wave equation can be made invariant, if in the moving frame, the time variable is changed together with the spatial variables. Since the wave equation was believed to describe electromagnetic phenomena in terms of potentials, the non-invariance of the former entailed the non-invariance of Maxwell's electrodynamics.

We present here the argument of noninvariance for the scalar potential, $\phi$,  for which a wave equation can be derived \emph{in vacuo}, namely
\begin{equation}
\phi_{tt}=c^{2}(\phi_{xx}+\phi_{yy}+\phi_{zz}),\label{eq:wave}
\end{equation}
where $c$ is the characteristic speed of linear waves. It is sufficient to consider only the one-dimensional case, when there is no dependence on $y$ and $z$. The question arises: ``Does Eq.~\eqref{eq:wave}  describe phenomena that are invariant when changing to a moving coordinate frame?''  To answer this question, one has to consider the moving frame:
\[x^{\prime}=x-vt, \quad t'=t \qquad  \text{or} \qquad  x=x^{\prime}+vt^{\prime}, \quad t=t',  \]
where $v$ is the velocity of the center of the moving coordinate frame.  Then the scalar potential in the moving frame is given by $\Phi(x^{\prime},t^{\prime})=\phi(x^{\prime}+vt^{\prime},t^{\prime})$, or, alternatively $\phi(x,t)=\Phi(x^{\prime}-vt^{\prime},t^{\prime})$. Upon expressing the partial derivatives of $\phi$ via the partial derivative of $\Phi$ we get
\begin{equation}
\phi_{tt}- c^{2}\phi_{xx} =
\frac{\partial^2\Phi}{\partial t'^2} - 2v \frac{\partial^2\Phi}{\partial t'\partial x'}- (c^2-v^2) \frac{\partial^2\Phi}{\partial x'^2},
\label{eq:wave_moving}
\end{equation}
which has a different form than Eq.\eqref{eq:wave}. This means that in the moving frame, different properties of the electromagnetic waves will be observed. Since nothing of this kind happens in the observations, the conclusion was that something is fundamentally wrong with the whole concept of physics.

One of the casualties of the perceived non-invariance of electrodynamics was the concept of a material medium that pervades the  space and serves as the transmitter of the electromagnetic waves. The luminiferous medium was believed to be a very rarefied ethereal substance (the `aether').  As it will be shown in the present work, the material continuum is not an ethereal substance although there is no reason to believe that it does not exist.  What happened with the dismissal of the aether was that the baby was thrown out with the bath water.  If one admits that space is not merely an empty `geometric' vessel for the physical processes, then one should realize that the time derivative in any kind of motion must be the  convective derivative (called in different contexts `material', `total',  or `substantial' derivative), 
\begin{equation}
 \frac{D}{Dt} = \DD{}{t} + v\DD{}{x}, \label{eq:material_deriv}
\end{equation}
not merely the partial time derivative.

If the convective derivatives are substituted \emph{in lieu}  of the local time derivatives, the paradoxical features of the model are instantly removed. It is interesting to note that the concept of convective derivative was around more than a century (Euler, \emph{circa} 1755) before Maxwell formulated his equations. Even more, Hertz \cite{Hertz} proposed to use the convective time derivative in Maxwell's equations  when dealing with the electromagnetic field \emph{in} material continua. The Maxwell--Hertz equations are called also `progressive wave equations' and are obviously translatory invariant because of the material derivatives involved. Apparently, this point was originally raised in \cite{Phipps}, as reported in \cite{Cantrell}, where the case for Galilean invariance of electrodynamics is forcefully argued.

The Hertz version of the equations of electrodynamics was not appreciated in full because the scientists did not think about space itself as being a material continuum (what we call these days `physical vacuum' \cite{TDLee,PhysVac}). The question was being raised of whether the progressive-wave equations can be construed to also hold \emph{in vacuo}.  Contrary to the tenets of the current orthodoxy, the answer is obviously in the affirmative, provided that one accepts the trivial fact that space (often called \emph{physical} vacuum) has to be a physical (material) continuum.  Unfortunately, Hertz's proposal did not attract much attention, and was discarded on the grounds that empty space is something different from a mechanical continuum. Thus, the electromagnetic field was assumed not to be a material substance, but rather something else.

As a `non-material' way out of the quandary of the perceived non-invariance, Voigt and Lorentz proposed to change the time in the moving frame. Their derivations differ from each other by the presence of the Lorentz factor, but both of them (and Larmor and Poincare  virtually simultaneously with them), espoused essentially the same idea that time was no longer an  absolute parameter, but could change from frame to frame, being thus influenced by the motion of the frame. What is nowadays called the Lorentz transformation,  is the following change of variables
\begin{gather}
x^{\prime}=\gamma(x-Vt), \  t'=\gamma(t -\frac{V}{c^2}x) \quad  \text{or} \quad  x=\gamma(x^{\prime}+Vt^{\prime}), \ t=\gamma(t'+\frac{V}{c^2}x'), \label{eq:LT}
\end{gather}
where $\gamma = [1-{v^2}/{c^2}]^{-\frac{1}{2}}$. It is straightforward to demonstrate that this change of variables leaves the linear wave equation invariant, namely
\begin{equation}
\phi_{tt}- c^{2}\phi_{xx} = \gamma^2\big( 1- V^2/c^2\big)\Big(
\frac{\partial^2{\Phi}}{\partial t'^2} -c^2 \frac{\partial^2{\Phi}}{\partial x'^2}\Big)
\equiv \Big(
\frac{\partial^2{\Phi}}{\partial t'^2}-c^2 \frac{\partial^2{\Phi}}{\partial x'^2}\Big).
\label{eq:wave_moving_Lorentz}
\end{equation}

The limited success of the Voigt--Larmor--Lorentz transformation is connected to the fact that it tacitly restores to some extent the convective derivative, i.e. it emulates the material invariance for  non-deformable frames in rectilinear motion (see \cite{Chri_FP}). Indeed, in 1D moving frame, the Lorentz transformation gives the  the following
\begin{equation}
 \DD{}{t} =  \gamma \DD{}{t'} + \gamma V\DD{}{x'}, \label{eq:Lorentz_1st_deriv}
\end{equation}
which resembles very much the convective derivative Eq.~\eqref{eq:material_deriv}, save the contraction factor, $\gamma$, and the fact that $V$ in Eq.~\eqref {eq:Lorentz_1st_deriv} is constant in space and time. Yet, Eq.~\eqref{eq:Lorentz_1st_deriv} is not equivalent to Eq.~\eqref{eq:material_deriv}, because the latter is valid for any local velocity (the frame can move non-inertially and even deform during the motion), while the former is true only for rectilinear motion of the frame. No one has been able to generalize the LT for accelerating frames, not to mention generally deforming frames. An interesting modification of LT to the case of uniformly accelerating frames can be found in \cite[Ch.18]{HsuHsu}, but the invariance of Maxwell equations is possible for this case only if the speed of light is a linear function of space.  This means that as of today, the `covariance' in the term `Lorentz covariance' is merely wishful thinking and is approximately true only in rectilinearly moving non-deforming frames. For this reason, the present author called the crippled invariance known as  `Lorentz covariance' the \emph{`Poor Man's Material Invariance'} \cite{Chri_FP}.  The point of the present work is that at this junction of time, the problem with the covariance has by no means been solved, and one has to continue the quest for the invariance of  electromagnetic phenomena.

Finally, is invariance possible without demoting the absolute statue of time? The answer is of course `no', if one looks at the original Maxwell equations.  However, the correctly rederived equations of electrodynamics are \emph{material} invariant which hints at the idea that the true covariance lies in the material invariant description (elements of which are called  `frame indifference' in mechanics of continua). Here we will be guided by the following obvious statement:
\begin{quote}
{\sf
 ``The material world must be material invariant'',}
\end{quote}
and will provide the mathematical technique needed for the theory of absolutivity, which emerges from the conjecture that there exists an absolute continuum. We introduce the concept of the absolute continuum as an elastic liquid (metacontinuum) and derive the model from the principles of rational continuum mechanics \cite{Chadwick,Marsden,Truesdell}.

\section{The Elastic Metacontinuum}\label{sec:metacont}

An absolute continuum is more naturally modeled in the frame of  material (Lagrange) coordinates $\bm{X}$. This approach is called the `referential description' (see, e.g. \cite{Chadwick,Marsden,Truesdell}). The connection of the geometric (Euler) coordinates $\bm{x}$ in the so-called `current configuration' is given by
\begin{equation}
\bm{x} = \bm{x}(\bm{X};t) \quad \text{or} \quad \bm{X} = \bm{X}(\bm{x};t). \label{eq:Euler}
\end{equation}
The first equation defines, in fact, the trajectory in the geometric space of a given point of the material continuum, as specified by its `address' $\bm{X}$ in the material space. Note that $\bm{x}$ are functions of time while $\bm{X}$ are independent of time. Hence the explicit dependence on time in  Eq.~\eqref{eq:Euler}$_2$ is needed to `cancel' the time dependence introduced through the geometric coordinates $\bm{x}$. It is convenient to denote the gradients in the referential and the current descriptions as
\begin{equation}
\mathrm{grad}  \stackrel{\mathrm{def}}{=} \frac{\partial}{\partial \bm{x}},
\quad \mathrm{Grad}  \stackrel{\mathrm{def}}{=}  \frac{\partial}{\partial \bm{X}}.
\end{equation}
We follow the standard notations which use capital letters for a differential operation (such as gradient or divergence) in the referential description (material coordinates $\bm{X}$), and a small letter for differentiation with respect to the spatial variables.

Without loss of the generality, one can assume that at the initial moment of time $t=0$ the material and geometric coordinates coincide, i.e.
\begin{equation}
 \bm{x}_0 \stackrel{\mathrm{def}}{=} \bm{x}(\bm{X};0) = \bm{X}.
\end{equation}
Then one can define the displacement of a point of the continuum as
\begin{equation}
\bm{U}(\bm{X};t) \stackrel{\mathrm{def}}{=} \bm{x} - \bm{x}_0 = \bm{x}(\bm{X};t) - \bm{X}.
\end{equation}
Acknowledging the connection between Lagrange and Euler coordinates, Eq.~\eqref{eq:Euler}, one can write the displacement equivalently as a function of the later, namely
\begin{equation}
 \bm{u}(\bm{x};t) = \bm{U}[\bm{X}(\bm{x};t);t],
\end{equation}
where we use a different case for the letter for the displacement in order to stress the point that the functional dependence is now different. Naturally, the numerical values of $\bm{u}$ are the same as $\bm{U}$ if the coordinates are related by the transformation formulas, Eq.~\eqref{eq:Euler}. The velocity of a material point is then the mere time derivative of the displacement when the latter is thought of as a function of the material coordinates and time. The velocity can be expressed as a function of the geometric coordinates, namely
\begin{equation}
\bm{V}(\bm{X};t) = \bm{U}_t(\bm{X};t),
\quad \text{or} \quad \bm{v}(\bm{x};t) = \bm{V}[\bm{X}(\bm{x};t);t],
\end{equation}


A most important characteristic is the deformation gradient
\begin{equation}
\mathbb{F} = \mathrm{Grad\>} \bm{x} \ \Big( F_{i\alpha} = \frac{\partial x_i}{\partial X_\alpha} \Big), \quad 
\mathbb{F}^{-1} = \mathrm{grad\>} \bm{X} \ \Big( F^{-1}_{\alpha i} = \frac{\partial X_\alpha}{\partial x_i}\Big),\quad  J = \mathrm{det}(\mathbb{F}),
\label{eq:defgrad}
\end{equation}
where we reserve the block-capital letters for tensors that are defined simultaneously in the referential and the current configuration.

The Cauchy balance law can be written as (see, e.g., \cite{Bland,Chadwick,Marsden,Truesdell})
\begin{equation}
\hat\mu \frac{D^2 \bm{U}}{D t^2}\equiv \hat\mu\DDD{\bm{U(\bm{X};t})}{t} = \mathrm{div} \sigma = J^{-1}\mathrm{Div} \mathfrak{S}= \frac{\hat \mu }{\mu}\mathrm{Div} \mathfrak{S},
\label{eq:lagrange_tensor}
\end{equation}
where $D/Dt$ denotes the material time derivative (which is merely the partial time derivative in the referential description), and $Div$ is the divergence with respect to the Lagrange coordinates. Here, $\mu$ is the density in the
material reference frame, while $\hat\mu$ is the density in the current description (Euler variables): The equation of continuity $ \mu/\hat\mu = J$ is acknowledged in the above derivation. The Cauchy stress tensor in the current description is denoted by $\sigma$. Respectively,  $\mathfrak{S}=J \mathbb{F}^{-1} \mathbb{\sigma},$ is the Piola-Kirchhoff stress tensor (sometimes called Lagrange stress tensor \cite{Bland}). We use the `Gothic' fonts for tensors defined solely in the referential description.
Now, Eq.~\eqref{eq:lagrange_tensor} can be rewritten as
\begin{equation}
\mu\DDD{\bm{U(\bm{X};t})}{t} = \mathrm{Div} \mathfrak{S},
\label{eq:lagrange_variables}
\end{equation}
which is the final form of the Cauchy equations of motion in the referential description (Lagrange variables).

The main concept developed in our previous works (see \cite{Chri_ISIS_1,Chri_ISIS_2,Chri_MATCOM09})  is that the luminiferous continuum is at rest and what is currently understood as particles and charges, are phase patterns in the continuum. As shown in the cited works, the notion of the soliton (or quasi-particle) explains the presence of particles and charges (matter) as the phase patterns on the \emph{3D hyper-surface}/\emph{metacontinuum}. This is the reason to call the luminiferous medium a `metacontinuum', in the sense that it is beyond the particles and charges and  is their progenitor.

\section{Linear Elasticity and Maxwell's Equations \emph{in vacuo}}\label{sec:Maxwell}

The linearized governing equations are valid only for infinitesimal deformations, when the referential and spatial descriptions have the same form, provided that $\bm{X}$ is thought of as being the same as $\bm{x}$. The linear constitutive relationship for an elastic body (in the spatial description for definiteness) reads
\begin{equation}
\sigma = (\lambda + \eta) (\mathrm{div} \bm{u}) + \frac{\eta}{2} (\mathrm{grad}\>  \bm{u} +  \mathrm{grad}\>  \bm{u}^T), \label{eq:elastic_constitut}
\end{equation}
where $\lambda$ and $\eta$ are the Lam\'e coefficients (see, e.g., \cite{LandauLif,Segel}).
The  above constitutive law substituted in Eq.~\eqref{eq:lagrange_tensor} yields the so-called Navier equations (see, e.g., Segel (1987), p.117) for the displacement vector $\bm{u}(\bm{x};t)$
\begin{equation}
\mu\DDD{\bm{u}}{t} = (\lambda +\eta)\nabla (\nabla\cdot\bm{u}) + \eta \nabla^2 \bm{u}
=(\lambda +2\eta)\nabla (\nabla\cdot\bm{u})- \eta\nabla\times\nabla \times\bm{u},
\label{eq:linearized}
\end{equation}
where we can use the `nabla' operator $\nabla$ because the equations are written in the current description.

The speeds of propagation of shear and compressional disturbances, are given respectively by
\begin{equation}
c = \sqrt{\eta/\mu}
\quad c_s = \sqrt{(2\eta + \lambda)/\mu}, \quad \text{and} \quad \delta = \frac{\eta}{2\eta + \lambda} ={ c^2}/{c_s^{2}}
\label{eq:speeds}
\end{equation}
is introduced for convenience. Since the light is understood to be a shear phenomenon, one can call $c$ the speed of light. Respectively $c_s$ is the speed of `sound' of the metacontinuum.

\subsection{Large Dilational Modulus and Incompressibility of Metacontinuum}
In a compressible elastic medium, both the shear and the dilational/ compressional waves should be observable. Since the groundbreaking works of Young and Fresnel, it was well established that electromagnetic waves (e.g., light) are a purely transverse (shear) phenomenon. This observation requires us to reduce the complexity of the model and to find a way to eliminate the term proportional to the dilational modulus $\lambda$. Cauchy assumed that $\lambda=0$ and ended up with the theory of so-called `volatile aether' (see \cite{Whittaker}). Upon a closer examination, we found that such an approach cannot explain Maxwell's equations.

Let us now assume that the dilational waves are not observable because the other extreme situation is at hand: $\lambda \gg \eta$ which is equivalent to  $\delta \ll 1$ or $\delta^{-1} \gg 1 $. It is convenient to rewrite Eq.~\eqref{eq:linearized} in terms of the speed of light, $c$, and parameter $\delta$, namely
\begin{equation}
\delta\left(c^{-2}\DDD{\bm{u}}{t} + \nabla\times\nabla \times\bm{u}\right)
=\nabla (\nabla\cdot\bm{u}),
\label{eq:linearized1}
\end{equation}
and to expand the density $\mu$,  displacements $\bm{u}$, and velocities $\bm{v}$  into an asymptotic power series with respect to $\delta$, namely
\begin{equation}
\mu = \mu_0 + \delta \mu_1 + \cdots \ ,\quad
\bm{u} = \bm{u}_0 + \delta \bm{u}_1 + \cdots \ ,\quad
\bm{v} = \bm{v}_0 + \delta \bm{v}_1 + \cdots \ .
\label{eq:21}
\end{equation}
Introducing (\ref{eq:21}) into (\ref{eq:linearized1}) and combining the terms with like powers we obtain for the first two terms
\begin{subequations}
\begin{align}
\nabla(\nabla \bm{\cdot} \bm{u}_0) & = 0 \>, \label{eq:22} \\
c^{-2} \DDD{\bm{u}_0}{t}
+ \nabla \times \nabla \times \bm{u}_0 &=
\nabla (\nabla \bm{\cdot} \bm{u}_1) \stackrel{\mathrm{def}}{=} - \frac{1}{\mu c^2} \nabla \phi\>,
\label{eq:222}
\end{align}
\end{subequations}
where $\phi$ is introduced for convenience and is the potential of the spherical part of the internal stresses. It has dimension $\mu c^2$ and plays the same role as the pressure in an incompressible medium: $\phi$  is an implicit function in Eq.~\eqref{eq:222} that provides the necessary degree of freedom  to enforce the satisfaction of the `incompressibility' condition, Eq.\eqref{eq:22}. The latter can also be rewritten as
\begin{equation}
 \nabla \bm{\cdot} \bm{u}_0 = const, \quad \Rightarrow \quad \nabla\cdot \bm{v}_0=0,
\label{eq:solen}
\end{equation}
which requires that the velocity field be solenoidal within the zeroth-order of approximation of the small parameter $\delta$. From now on, the subscript `$0$' will be omitted form the variables without fear of confusion.

Eq.~\eqref{eq:222} can be rewritten as
\begin{equation}
\mu \DD{\bm{v}}{t} = -\nabla\phi + \bm{t}, \quad \bm{t} \stackrel{\mathrm{def}}{=} \nabla\cdot \sigma=  - \eta\nabla \times \nabla \times \bm{u},\\
\label{eq:momentum}
\end{equation}
where $\bm{v}\equiv\bm{u}_t$ is the velocity vector and $\bm{t}$ is the tangential part of the stress vector in the metacontinuum.  

\subsection{Maxwell's Equations as Corollaries}
The system, Eq.~\eqref{eq:momentum}, lends itself to a far-reaching analogy if one  names the negative stress vector the `electric force,' and defines the  `magnetic field,'  $\bm{H}$, as the vorticity in the metacontinuum, namely
\begin{equation}\bm{E} \stackrel{\rm def}{=} -\bm{t}
=  \eta\nabla\times ( \nabla \times \bm{u}),
\qquad \bm{H} \stackrel{\rm def}{=} \nabla\! \times\! \bm{v}, \quad  \bm{B} = \mu \bm{H}. \label{eq:elmag_definitions}
\end{equation}
where the magnetic induction is defined the usual way for the electrodynamics \emph{in vacuo} and the density of the metacontinuum is the  `magnetic permeability'.

Now taking the $curl$ of Eq.~\eqref{eq:momentum} and acknowledging the notations \eqref{eq:elmag_definitions}, we get Faraday's law
\begin{equation}
\nabla \times \bm{E} =  -\DD{\bm{B}}{t}.
\label{eq:faraday}
\end{equation}

On the other hand, taking the time derivative of Eq.~\eqref{eq:elmag_definitions}$_1$ we get
\begin{equation}
\frac{1}{\eta}\DD{\bm{E}}{t}  =  \nabla \times (\nabla \times
\bm{v}) \equiv \nabla \times \bm{H} ,
\label{eq:rheology_law}
\end{equation}
or in terms of magnetic induction
\begin{equation}
\frac{\mu}{\eta}\DD{\bm{E}}{t} = \nabla \times \bm{B}
\quad \Rightarrow \quad
\DD{\bm{E}}{t}  = c^2 \nabla \times \bm{B} ,
\label{eq:second_Max_proto}
\end{equation}
which is exactly the second of Maxwell's equations (or Maxwell-Ampere's equation). The interpretation of the shear elastic modulus is that its inverse defines the electric permittivity in electrodynamics.

\subsection{The Metacontinuum: Liquid or Solid?}

The previous subsection unequivocally shows that the `field' described by Maxwell's equations is equivalent to an elastic material space (the metacontinuum). To author's knowledge the connection of Maxwell's equation to the equations of elastic medium was first established in \cite{Chri_WS,Chri_CMDS8,Chri_annuary}. Apparently, similar derivations were proposed independently in\cite{Dmitriev,Karlsen}, and just recently in \cite{Wang}. The common trait of those papers is that the metacontinuum is considered as an elastic \emph{solid}.  In such a model, no \emph{stationary} magnetic fields can  exist, since no steady velocities are possible for a solid continuum without discontinuities. Despite the fact that a liquid rheology is mentioned in \cite{Wang}, the model used for the derivation of electrodynamics is the elastic solid. In the notations of \cite{Wang}, stationary electric fields would be impossible for the same reason mentioned above.

In the present paper we make the next decisive step in developing the model: we consider elastic liquid \emph{in lieu} of an elastic solid. This means that for shear deformations, the metacontinuum must be an elastic \emph{liquid} for which the time derivative of the deviator  stress tensor is related to the deviator rate of deformation tensor $\chi$, namely
\begin{equation}
\sigma_t = \eta \chi, \quad  \chi=   \tfrac{1}{2}(\nabla \bm{v} +\nabla \bm{v}^T), \label{eq:tensor_rheology}
\end{equation}
This rheology can be rewritten for the negative stress vector $\bm{E}=- \nabla\cdot \sigma$ and deformation vector, $\bm{d} = \nabla \cdot \chi(=- \nabla \times \nabla\times \bm{v}$ for $ \nabla\cdot\bm{v} = 0$), since both of these vectors are  the divergences of the respective tensors involved in the elastic rheology, Eq.~\eqref{eq:tensor_rheology}.  Then
\begin{equation}
\bm{E}_t = -\eta \bm{d} = \eta \nabla \times \nabla \times\bm{v}
\quad \text{or} \quad \tau \bm{E}_t = \zeta \nabla \times \nabla \times\bm{v} ,
\end{equation}
where $\zeta$ is called `elastic viscosity' \cite{Joseph}. In the alternative version    $\tau$ is the relaxation time of the stresses, and the apparent elastic shear modulus is given by $\eta = \zeta\tau^{-1}$. 
Note that the above elastic-liquid rheology concerns just the shear deformations.  For compressional/dilational motions, the metacontinuum can still behave as a virtually incompressible solid. A more general formulation of the shear part of the constitutive relation would be the viscoelastic liquid
\begin{equation}
\tau \bm{E}_t + \kappa \bm{E} = \zeta \nabla \times \nabla \times \bm{v} \quad \text{or} \quad
\bm{E}_t + s \bm{E} = \eta \nabla \times \nabla \times \bm{v},
\label{eq:constitutive_solid_liquid}
\end{equation}
where $s= \kappa\tau^{-1}$ can be called `conductivity' of the viscoelastic liquid. Note that in \cite{Joseph}, the conductivity is set to unity, which precludes treating purely elastic liquids (in the sense of non-viscous). We prefer to keep the flexibility offered by the introduction of the coefficient $\kappa$. Setting the appropriate terminology is a complicated task, because, when $\kappa\ne 0$, then $\zeta$ does indeed have a meaning of viscosity, while for $\kappa=0$, it loses its independent meaning and enters the picture through the  coefficient of apparent shear elasticity $\eta$. 

The case $\kappa\ne 0$ can be called viscoelastic rheology, but it does \emph{not} lead to a Navier-Stokes liquid with additional elasticity because there is no retardation term that involves the time derivative of the deformation tensor/vector. In this sense, adding the conductivity does not introduce dispersive dissipation, but rather a linear attenuation parameterized by the conductivity coefficient. For the effects connected with the attenuation/conductivity we refer the reader to \cite{Chri_MATCOM07} and \cite{Harmuth}. The constitutive relationship given in Eq.~\eqref{eq:constitutive_solid_liquid} can be interpreted as stipulating Ohm's law for \emph{vacuo}. Although, such kind of stipulation is made in main texts (e.g., \cite{Joos}), the cause of Ohm's law in matter can be the thermal fluctuations of the atoms that obstruct the free passage the charges through a conductor. Clearly, a more in-depth argument is needed to justify having Ohm's law in \emph{vacuo}, which goes beyond the scope of the present paper.

In a nutshell, the effect of a very small conductivity of the metacontinuum can lead to reconsidering the luminosity of the nearby galaxies, which would  increase their estimated mass, and hence could alleviate the need of introducing the concept of so-called `dark matter'. Conversely, the fact that the mass of Universe appears to be bigger than the one estimated on the basis of luminosity, speaks in favor of having, albeit small, but nontrivial coefficient of conductivity.

\subsection{Effect of Compressibility (Dark Energy?)}

There are no conceptual difficulties in extending the above described model of electrodynamics to include a slight compressibility of the metacontinuum. Such a generalization raises the question about the speed of the compressional waves (`sound') of the metacontinuum. This speed is expected to be much greater ($\delta^{-1}$ times greater) than the speed  of the shear waves, $c$. However, the investigation of the quantitative effects of a slight compressibility goes beyond the scope of the present paper.  In order to avoid ambiguous terminology we will not use the term \emph{sound} for the compression waves in the metacontinuum. Rather we will borrow a coinage from the ancient school of Stoa (see \cite{Sambursky}) and will call the compressional/dilationlal motions the \emph{pneuma}.

One obvious implication of the existence of waves of a different kind than the shear (electromagnetic) waves is that, in fact, there is more energy in the physical vacuum (metacontinuum) than what is detected from electromagnetic radiation. The \emph{pneuma} waves are orthogonal to the shear waves and are not detectable by standard devices based on  electromagnetic interactions. They perfectly fit the bill of what is currently called `dark energy' (see \cite{HutererTurner}).  Over the last couple of decades, observations of supernovae have shown that the supposed expansion of the Universe is accelerating \cite{Perlmutter,Riess}. One way to quantitatively model this acceleration is to `play' with the cosmological constant  (\cite{PeeblesRatra}). Yet, a physical (mechanical) cause for variability of the cosmological constant is still lacking. Recently, a consensus has emerged around the conjecture that the Universe is expanding faster than expected because of the presence of (what some have called `negative') pressure that acts through  space and pushes the matter apart.

It is important to investigate the interaction between the shear and longitudinal vibrations. For instance, the transfer of energy under certain conditions from the longitudinal (electromagnetically undetectable) mode to the shear (electromagnetically detectable) mode may turn out to be the cause of unexplained sources of electromagnetic radiation (e.g., gamma-rays bursts). In the linear approximation the \emph{pneuma} waves and the electromagnetic waves do not couple, so the next stage of the present theory will be to include the effects of finite elasticity in the constitutive relations.


In concluding this section, we mention that Eqs.~\eqref{eq:momentum},\eqref{eq:rheology_law}, and \eqref{eq:constitutive_solid_liquid} govern the propagation of transverse (shear) waves and accomplishes the goal of Cauchy who attempted to explain light waves as shear waves of a material medium. A very dense jello (or a pine pitch) reacts to high-frequency shear undulations essentially in the same manner as a polymeric material.

\section{Localized Torsional Dislocations in the Metacontinuum}\label{sec:twistons}

In order to understand the interplay between the absolutivity of the metacontinuum and the relativity of the rectilinear motion, as discussed by Einstein \cite[Ch.7]{Einstein2},  we consider a $2D$ case with only two nonzero velocity components, $v_x(x,y)$ and $v_y(x,y)$  and no motion of the \emph{material} frame, i.e., no predominant velocity component of the material points of space/metacontinuun.  In 2D, we can introduce a stream function:  \( v_x = \partial_y{\psi}\), \(v_y = -\partial_x{\psi}\), and from the linearized governing equations \eqref{eq:linearized} the following wave equation can be derived
\begin{equation}
{\partial}_{tt}\psi - c^2 (\partial_{xx} + \partial_{yy})\psi = 0.
\label{eq:linearized_wave}
\end{equation}

\begin{figure}[h!]
\centerline{\includegraphics[width=0.82\textwidth]{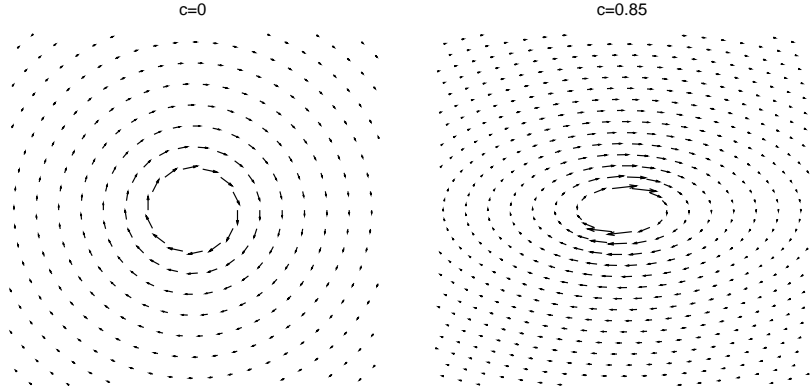}}
\caption{A localized gradient phase pattern (twiston) in two dimensions. Left panel: pattern at rest. Right panel: Pattern propagating with phase velocity $(w_1,w_2)$.}\label{fig:tLorentz_twistons}
\end{figure}

The last equation has a stationary solution with polar symmetry: $\psi = \ln(r)$, where $r=\sqrt{x^2+y^2}$. For the velocity  components, one has the following expressions  \(v_x = {y}/{r^2},  v_y =-{x}/{r^2}\). This solution represents the well known potential vortex (in the left panel of Fig.~\ref{fig:tLorentz_twistons}) which has a nontrivial circulation (topological charge). This suggests that the localized torsional deformations can be interpreted as the charges. In what follows, we term this kind of a localized wave pattern the `twiston.'

Now, we examine the situation when a localized solution of the above type \emph{propagates} with phase velocity $\bm{w}=(0,w)$.
There are no essential difficulties in considering a more general moving frame $\bm{w}=(w_1,w_2)$, but we resort to the simple translation along the $y$-axis for the sake of simplicity.
Note that even for a large phase speed of the pattern, $|\bm{w}| \lessapprox c$, the actual magnitude of the velocity of the material points of the metacontinuum is still very small, and thus one can use the linearized equation, Eq.~\eqref{eq:linearized_wave}. Then, in the  moving frame $\hat x=x$, $\hat y=y-wt$, one gets from Eq.~\eqref{eq:linearized_wave} that
\(
{\psi}_{\hat x\hat x}+(1-w^2){\psi}_{\hat y\hat y}=0,
\)
which possesses a solution of the type
\begin{equation}
\psi = \ln{z}, \ \ z = \sqrt{{\hat x^2}+{\hat y^2}(1-w^2/c^2)^{-1}}.
\label{eq:moving_pattern}
\end{equation}

We observe that the analytic form of the solution is the same, but for a different radial-like coordinate, $z$. The lines of constant $z$ are ellipses, i.e. one is faced with a `twiston' whose streamlines are ellipses. This means that the moving  phase pattern, Eq.~\eqref{eq:moving_pattern}, undergoes contraction in the direction of motion proportional to the Lorentz factor. This situation is depicted in the right panel of Fig.~\ref{fig:tLorentz_twistons}.  Since, the potential of interaction of two localized waves (see \cite{ChriChri_PLA}) for a similar derivation in the case of $sine$-Gordon equation) depends on the asymptotic behavior of their `tails,' the fact that the charges are shortened in the direction of motions will lead to shortening of the distances between them in the same direction. This means that an assemblage of charges (i.e., a body) will be shortened in the direction of motion by the Lorentz factor.

Thus, the relativity of rectilinear motion of phase patterns and the Lorentz contraction are manifestations of the \emph{absolutivity of space}. The patterns \emph{propagate over}, rather than \emph{move through}, the \emph{metacontinuum} and do not create an `aether wind,' i.e. the rectilinear translation of phase patterns does not contradict the principle of material invariance since the latter holds for the points of the material metacontinuum. The above derivations have a limited quantitative importance due to  their 2D nature, but their qualitative significance is very important because they show the interplay between the notions of absolute continuum and relative rectilinear motion.

An important caveat here is connected with the fact that the solution \eqref{eq:moving_pattern} is singular at the center of the coordinate system. The concept of a singularity is not one that is easy to reconcile with common sense. A similar singularity was encountered in the famous Schwarzschild's solution for the `black holes'. Einstein considered the Schwarzschild singularity as an  artifact. Similarly, in the case of the \emph{twiston}, Einstein's objection is equally valid. In general theory of relativity the investigators are not bound by a mechanical construct and can freely speculate about singularities. However, here we must provide, at least, an outline of the way out of this situation.

As shown in \cite{Chri_ISIS_1}, the improper behavior can be mitigated if  higher-order derivatives are present in the models. In elasticity, this is called  `gradient elasticity' (see the original work of Mindlin (1964), and the illuminating presentation in \cite{SharmaGanti}, from the point of view of a general Lagrangian formalism for the higher-gradient elasticity). The issue of higher derivatives involved in viscous liquids is thoroughly elucidated in \cite{JordanPuri1} and \cite{QuintaStraughan} in the context of dipolar and Green and Naghdi's  liquids \cite{GreenNaghdi}. Here is to be specially stressed that Boussinesq \cite{Boussinesq} did include dispersion in his aether model in order to  explain the dispersion of light, but his higher-grade elasticity was connected with the rotational inertia (mixed time/spatial fourth order derivative). Such kinds of dispersion cannot help in alleviating the problem of singularity.

Now, Eq.(3.7) of \cite{SharmaGanti}, Eq.(2.1) of \cite{QuintaStraughan}, and Eq.(2.10) of \cite{JordanPuri1}, all of them give the following generic equation for the stationary \emph{twiston} with the higher-gradients acknowledged:
\begin{equation}
\Delta \bm{v} - \chi \Delta\Delta  \bm{v} =0, \quad \chi = \hat \eta/\eta,
\label{eq:disperse}
\end{equation}
where $\hat \eta$ is the coefficient of the higher-gradient elasticity (or viscosity).  In terms of the above defined stream function, we can reduce eq.\eqref{eq:disperse} for the case of radial symmetry to the following
\begin{equation}
\frac{1}{r}\frac{d}{dr}r\frac{d}{dr}\big[ \chi \frac{1}{r}\frac{d}{dr}r\frac{d}{dr} \psi - \psi \big]= 0,
\label{eq:bessel}\end{equation}
which is shown in \cite{Chri_ISIS_1} to possess the following solution
\begin{equation}
 \psi = K_0(r/\sqrt{\chi}) +\ln (r/\sqrt{\chi}).
\end{equation}
The last equation does not have a singularity at the origin since $K_0(r) \propto -\ln\> r$ for  small $r$.  This means that the higher-gradient effects have to be considered in a more practical model of the metacontinuum. Since the higher-gradient terms play the role of dispersion, the question arises if there are some observations suggesting that dispersion could be present in the metacontinuum. The best candidate for a dispersion related effect is the cosmological redshift. The present author showed in \cite{Chri_annuary} that the higher gradient can actually lead to stretching of the internal length of an elastic  solitary wave. Alternatively, it is found in \cite{Chri_WM08} that the dissipation can make a narrow wave packet shift to the red (reduce its central wave number) without changing its apodization function much. This means that in a metacontinuum with higher-gradient elasticity or viscosity, a cosmological redshift will necessarily be present.

The detailed description of the higher-gradient generalizations goes beyond the scope of the present work.

\section{Euler Variables and Material Invariance}

As already discussed in Section~\ref{sec:dead_end}, the laws of physics (including the laws of continuum physics) must have the same form in any reference frame (coordinate system).  Unlike what is called `Lorentz Covariance', the laws in referential description are \emph{frame indifferent}, i.e. they are truly covariant.  However, the experimental measurements are always connected with a current frame in the geometric space. This means that an observational frame comprised by quasi-particles (phase patterns) cannot detect the material variables, but rather can merely measure their counterparts in the current (geometric) frame. This is a typical situation in mechanics of continuous media where the reference configuration is often not related to any measurable frame. For this reason we need to reformulate the model from Section~\ref{sec:metacont} in the current description making use of Euler variables. This is the objective of the present section.

In terms of the velocity vector, the Cauchy balance, Eq.~\eqref{eq:lagrange_tensor}, can be rewritten as follows:
\begin{equation}
\rho \frac{D  \bm{v}}{D t} \stackrel{\mathrm{def}}{=} \mu \DD{\bm{v}}{t} + \mu \bm{v}\cdot \nabla \bm{v} = \mathrm{div} \sigma = -\nabla \phi - \bm{E},
\label{eq:Cauchy}
\end{equation}
where $D/Dt$ stands for the material derivative (called `convective' or `total' derivative in the current configuration). Remember that in the referential configuration, it is just the partial time derivative. Note also that for an incompressible metacontinuum the density is the same constant in both the referential and spatial descriptions and we denoted it by $\mu$. Also, we use the above defined negative stress vector related to the deviatoric part of the stress tensor, which we have called the `electric field', $\bm{E}$. The Cauchy balance, Eq.~\eqref{eq:Cauchy} can be rewritten in the so-called `Gromeka-Lamb form' \cite{Sedov}
\begin{equation}
\mu \DD{\bm{v}}{t} - \mu\bm{v}\times (\nabla \times \> \bm{v} )= -\nabla\big(\phi + \frac{\bm{v}^2}{2}\big) - \bm{E}, \label{eq:Lamb}
\end{equation}

Now, taking the \emph{curl} of Eq.~\eqref{eq:Lamb}, and using our definitions Eq.~\eqref{eq:elmag_definitions}, we get:
\begin{equation}
\nabla \times\>[ \bm{E}- \bm{v} \times  \bm{B}] = - \DD{\bm{B}}{t} ,
\label{eq:Faraday_Lorentz}
\end{equation}
which is the Faraday law with an additional force that  is exerted by a moving magnetic field in each point of the medium due to the convective part of the acceleration at that point. The reaction to this force is the force acting on a moving point (charge), $\bm{v}\times \bm{B}$, known as the `Lorentz force'. In other words, the material invariant Faraday's law automatically accounts for a physical mechanism that can cause the Lorentz force. This is a very important result, because it tells us that the Lorentz force is not an additional, empirically observed force that has to be grafted on the Maxwell model in order to  complete the electrodynamics, but is connected to the material time derivative, namely to its convective part. 

Under the incompressibility condition,  Eq.~\eqref{eq:Faraday_Lorentz} can be recast to
\begin{equation}
\DD{\bm{B}}{t} + \bm{v}\cdot \nabla \bm{B} -\bm{B}\cdot\nabla \bm{v} = - \nabla \times\>  \bm{E},
\label{eq:Faraday_Hertz}
\end{equation}
which we can call the `Hertz form' of the Faraday-Lorentz law. Note the presence of the third term on the left-hand side. It is not in Hertz's formulation, and it has been erroneously omitted in \cite{Chri_FP,Chri_MATCOM07}, too. Note that Eq.~\eqref{eq:Faraday_Hertz} does not give any special advantage over Eq.~\eqref{eq:Faraday_Lorentz}. The form Eq.~\eqref{eq:Faraday_Hertz} shows that one cannot just add the convective part of the derivative to Faraday's law to make it frame indifferent, because magnetic field is not a primary variable. Rather it is the $curl$ of the velocity vector, $\bm{v}$. The frame indifference for the primary variable gives as a corollary Eq.~\eqref{eq:Faraday_Hertz}.

As has already been shown in Section~\ref{sec:Maxwell}, the second of Maxwell's equations is in actuality the constitutive relation for the viscoelastic rheology of the metacontinuum, if written in terms of the stress and strain vectors. The concept of frame indifference (general covariance of the system) requires that the partial derivaive of the stress variable (in our case the stress vector $\bm{t}=-\bm{E}$) in Eq.~\eqref{eq:constitutive_solid_liquid} be replaced by the appropriate invariant rate. It is argued in \cite{Chri_FP} that the pertinent invariant rate is the so-called Oldroyd's upper-convected derivative \cite{Oldroyd}, namely
\begin{equation}
\frac{D \bm{E}}{D t} \stackrel{\mathrm{def}}{=}\DD{\bm{E}}{t} + \bm{v}\cdot \nabla \bm{E}- \bm{E} \cdot \nabla \bm{v} + (\nabla\cdot\bm{v}) \bm{E} ,
\label{eq:elfield_Oldroyd}
\end{equation}

Now using the vector identity (see, e.g. \cite[pg.180]{BoriTarap})
\begin{equation} \nabla \times (\bm{v} \times \bm{E}) = \bm{E}\cdot \nabla \bm{v} - \bm{v}\cdot \nabla \bm{E} + \bm{v} (\nabla\cdot \bm{E}) - \bm{E} (\nabla \cdot \bm{v}),
\label{eq:identity}
\end{equation}
the following generalization of the second of Maxwell's equation is obtained
\begin{equation}
\DD{\bm{E}}{t}  - \underbrace{\nabla \times (\bm{v} \times \bm{E})} + \underline{s\bm{E}} = - \widehat{\bm{v} (\nabla \cdot \bm{E})} + \widehat{\underbrace{ c^2 \nabla \times  \bm{B}}}.
 \label{eq:Max_Ampere}
\end{equation}

The terms with hats over them clearly give as a corollary the Ampere law upon introducing the definition of a charge
\begin{equation}
\rho \stackrel{\mathrm{def}}{=} \nabla\cdot \bm{E}, \label{eq:charge_def}
\end{equation}
whose meaning is now extended to electromagnetism \emph{in vacuo}. Namely, the `chargedness', $\rho$, of the free space is defined as the divergence of the electric field in the point. In order not to confuse this property of the field \emph{in vacuo} with the localized pattern called, say an electron or a proton (see Section~\ref{sec:twistons}), we call the above defined function $\rho$ the \emph{metacharge}.  For the latter, a continuity equation is readily derived upon applying the operation $div$ to Eq.~\eqref{eq:Max_Ampere}, namely
 \begin{equation}
 \DD{\rho}{t} + \nabla\cdot(\rho\bm{v} ) +s \rho =0, 
 \label{eq:charge_continu}
  \end{equation}
which, for the case $s=0$, is the standard continuity equation for charge. Here is to be mentioned that Eq.~\eqref{eq:charge_continu} was derived in \cite{Chri_FP} directly from the Oldroyd form, Eq.~\eqref{eq:elfield_Oldroyd}, but the derivation here is much more straightforward because of the application of the identity Eq.~\eqref{eq:identity}. We shall refer to Eq.~\eqref{eq:charge_continu} as the continuity equation for the \emph{metacharge}.

The terms with the underbraces in Eq.~\eqref{eq:Max_Ampere} are related to the Biot-Savart law, in the sense that they can be obtained after the operation $curl$ is applied to the latter. The sign of the term on the left-hand side has a sign opposite to the ubiquitous Biot-Savart law. Similarly to the above outlined argument for the Lorentz force, the sign is different because it refers to the electric field created at a point of the metacontinuum by the presence of a magnetic field. If one considers the electric field  as produced by a moving current, the velocity changes it sign, and the Biot--Savart law holds in its standard form. It is interesting to note (see \cite[P.498]{Griffiths}) that in the transformation of Maxwell equations under Lorentz transformation, the terms appear with the same signs as in Eq.~\eqref{eq:Max_Ampere}. 

A remarkable feature of Eq.~\eqref{eq:Max_Ampere} is that it includes both Oersted--Ampere's laws and Biot--Savart laws. An idea was floated by Whittaker \cite[pp.85--88,vol.I]{Whittaker} that both these laws may actually follow from a single law, similar to what is presented in Eq.~\eqref{eq:Max_Ampere}. There is still ongoing discussion in the literature about whether these two laws are identical (see, e.g.\cite{Christodoulides,Jolly}) or they are independent (see, e.g. \cite{Graneau}). Our results seem to favor Whittaker's  original idea that both of the laws have to present in the correct formulation. 

The important conclusion form the frame-indifferent formulation of the displacement current is that similarly to Lorentz-force law, the convective/ convected terms are related to phenomena that is embodied in Ampere's and Biot--Savart's laws, providing thus their possible unification with Maxwell's model. All three electromagnetic-force laws are manifestations of the inertial forces in the metacontinuum.

Now, Eqs.~\eqref{eq:Faraday_Lorentz},\eqref{eq:Max_Ampere} form the system that is to replace the first two of Maxwell's equations (the dynamic equations). Following the tradition, one can add also the obvious corollary of the fact that the magnetic field is the $curl$ of another vector (velocity), namely,
\begin{equation}
\nabla \cdot \bm{B}= 0.
\label{eq:B_diverge}
\end{equation} 

The above derived system generalizes the Hertz program from 1890 and \emph{rigorously} fulfills the requirements for `General Covariance,' because this system is frame-indifferent, i.e., it is invariant when changing to another coordinate frame that can accelerate and can even deform. A very limiting case of the frame indifference is the Galilean invariance.  

The frame indifferent model of the luminiferous continuum (called here the metacontinuum), succeeds to unify in a single nexus all known information about electromagnetism: Faraday's law, displacement-current law, Lorentz-force law, Oersted--Ampere's law, and Biot--Savart law. It is a significant step forward from Maxwell's model, in which only the first two were explained by the equations themselves, and the latter three appeared as additional empirically observed relation between the main characteristics of the field. Since these new terms are valid \emph{in vacuo}, they are clearly the progenitors of the respective phenomena in moving media.

\section{Detection of the Absolute Continuum}

 The celebrated experiment of Michelson and Morley (MME) \cite{Michelson1,MiMor} turned a nil result, which was construed as an indication that there exists no absolute medium with respect to which one can define a relative motion. This conclusion is hailed as one of the most important conceptual achievements of modern physics, because it led to a precipitous paradigm shift:  the mechanistic view of the nineteen century was replaced by a more abstract world view.  Yet, everybody seems to agree that, at least locally, one can consider a preferred frame.  A good overview of the situation and the challenges that the accepted point of view is still facing can be found in \cite{ConsCost}.

The problem with the above conclusion is that it presents a fallacy consisting of using an argument that is supposed to prove one proposition but succeeds only in proving a different one. The fallacy here is that the conclusion is overreaching. The only rigorous conclusion which can be drawn here is that the absolute continuum cannot be detected by this \emph{particular} experiment.  This, more limited, conclusion satisfies also Occam's razor.  It  was the guideline in the proposition of Fitzgerald (see, \cite[p.749]{Lodge_aberrat}) and Lorentz \cite{Lorentz92} who explained the nil effect by the possible contraction of the lengths in the direction of motion. There is no justification whatsoever to ``throw out the baby with the bath water'', and conclude that there is no absolute medium. The Lorentz--Fitzgerald contraction has been splendidly confirmed in all major experimental tests, and can be considered now as one of the most important discoveries in physics. Accepting the contraction assumption instantly rendered superfluous the dismissal of the absolute medium. 

The above outlined logical fallacy in the interpretation of MME  does not necessarily invalidate the assumptions that led to relativity, since one of the explanations of the nil effect is \emph{indeed} the possibility that an absolute continuum  does not actually exist, or the continuum is not absolute, but can be entrained by the moving matter. Yet, it is scientifically unsatisfactory to adopt  such a far-reaching conclusion without exhausting the possibilities to measure the relative speed.  This means that all effort must be made to create an experimental set-up that can demonstrate, unequivocally, the relative speed in a laboratory setting. It is really not that important if we call the preferred frame LSR or absolute continuum. What is more important is to measure the quantitative value of the relative speed.

Another celebrated experiment which is believed to have unequivocally supported the theory of relativity was performed by Ives and Stilwell (ISE). It was understood that the change of frequency of moving atoms in a cathode tube is due only to time dilation. This is another overreaching conclusion, because there is no one-to-one correspondence between a frequency change and a time dilation. The change can occur for various reasons, and one of them is discussed here.

\subsection{Time Dilation or Change of Frequency due to Relative Motion? }

\subsubsection{Emission of Light by Moving Atoms}

A photon is emitted/absorbed when an electron jumps from one orbit to another. If the atom is at rest with respect to the absolute continuum, then  for the particular electron jump, the frequency is defined by the distance, between the orbits. If so, the spatial wave number, $\nu$ of an emitted photon is governed by the Bohr--Rydberg law
\begin{equation}
\nu = R \Big(\frac{1}{n_f^2} -\frac{1}{n_i^2}\Big),
\end{equation}
where  $n_f$, $n_i$ are the numbers of the respective orbits between which the electron jumps. Respectively, 
\begin{equation}
R = \frac{2\pi^2 e^4m}{ch^3}\mathrm{cm}^{-1} \approx 109737 \mathrm{cm}^{-1},
\end{equation}
is the Rydberg number (see e.g. \cite{Joos}). Its inverse, $d \propto R^{-1}$,
defines the diameter of the lowest orbit.

\begin{figure}[h]
\centerline{\includegraphics[width=0.55\textwidth]{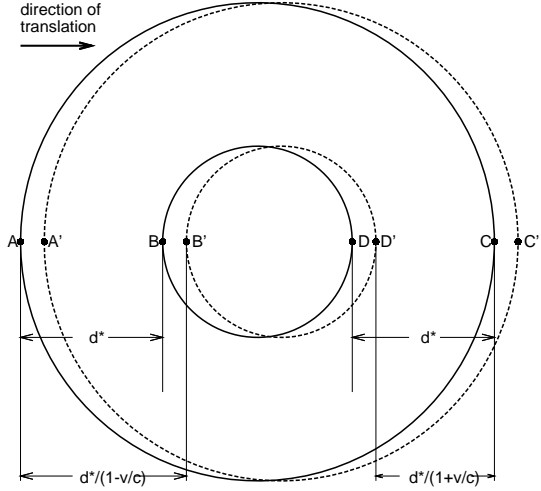}}
\caption{Bohr-Rydberg emission model for a moving atom.}\label{fig:Rydberg_Lorentz}
\end{figure}

The question now is how the emitted frequency is influenced by the fact that the atom is moving with respect to the supposed absolute medium. We can consider the atom as a translating phase pattern, whose dimensions are shortened in the direction of motion. This means that the orbits become curves on ellipsoids whose shortest axes are proportional to the inverse of the Lorentz factor, i.e.,  the smallest value of the characteristic distance becomes $d^*=d\sqrt{1-w^2/c^2}$, where $w$ is the speed of the relative motion of the atom with respect to the  absolute continuum  (see Fig.~\ref{fig:Rydberg_Lorentz}).

The important thing to understand here is that the photon is emitted by the atom, but detaches from it and lives a life on its own as a shear wave  in the absolute continuum.  A cross-section of an atom in motion is presented in Fig.~\ref{fig:Rydberg_Lorentz}. The solid  lines show the orbits at the moment $t$, while the dashed lines are the orbits for $t+T$, where $T$ is the time needed for the photon to traverse the distance between the two orbits. Note that the orbits intersect, but it does not mean that the electrons will collide, because they are at different positions on the orbits at different moments of time. Because of the Lorentz contraction, the orbits are actually curves that lie on ellipsoids.  It is natural to expect that a photon will jump between the closest periapsides of the two orbits under consideration. It makes an important difference if the photon is emitted alongside or against the motion of the atom. 

Let us begin with the case when the photon is emitted in the direction of translation (left part of Fig.~\ref{fig:Rydberg_Lorentz}). The actual length over which the photon is emitted is not $d$, and not even $d^*=\overline{AB}=\overline{A'B'}$, but the length $\overline{AB'}$.  At moment $t$, the photon initiates its jump at the point $A$ which is the periapsis of the lower orbit. At  $t+T$, the photon ends the jump at point $B'$ which is the periapsis of the outer orbit. Here  $T$ is the time needed for a photon to traverse the distance between points $A$ and $B'$, i.e., $cT= \overline{AB'}$. Then $\overline{BB'}= vT$, where, $v$ is the speed of the center of atom. Then 
\( cT =\overline{AB'} = \overline{AB} + \overline{BB'} = d^* +wT
\), 
which gives
\begin{equation}
T\big(1-\frac{w}{c}\big) =  \frac{d}{c}\sqrt{1 - \frac{w^2}{c^2}}
\  \Rightarrow  \  
 T = \frac{d}{c}  \frac{\sqrt{1 + w/c}}{\sqrt{1 -  w/c}}.
\end{equation}
Now, for the frequency, $\omega_m$, of the photon emitted from a moving atom in the direction of translation, we get
\begin{equation}
\omega_m = \frac{2\pi}{T} = \frac{2c\pi}{d}  \frac{\sqrt{1 - w/c}}{\sqrt{1 + w/c}},
\label{eq:relat_Doppler}
\end{equation}
where $\omega=2c\pi/d$ is the `rest' frequency of the respective spectral line that would have been emitted if the atom were in complete rest with respect to the absolute continuum. It is important to point out here  that 
Eq.~\eqref{eq:relat_Doppler} has the same form as what is called the `relativistic Doppler effect' (see, e.g. \cite{Gill}). The above described frequency change, can be called the `intrinsic Doppler effect' of the emission from moving atoms. 
Eq.~\eqref{eq:relat_Doppler}  is in fact a demonstration of the absolutivity of space, and does not require relativistic arguments to be derived.  The situation is very similar to the case with Lorentz contraction which was shown in Section~\ref{sec:twistons} to be easily explained by considering the charges as phase patterns which propagate  \emph{over} an absolute continuum. 

For completeness, we also consider the case when an electron jumps from point $C$ of the higher orbit backwards to the point $D'$ which is the point reached by the periapsis $D$ of the lowere orbit after $T$ units of time. Now
\( cT =\overline{CD'} = \overline{CD} - \overline{DD'} = d^* - wT\),  
which gives
\begin{equation}
 T = \frac{d}{c}  \frac{\sqrt{1 - w/c}}{\sqrt{1 + w/c}}
 \  \Rightarrow  \  
 \omega_m = \frac{2\pi}{T} = \frac{2c\pi}{d}  \frac{\sqrt{1 + w/c}}{\sqrt{1 - w/c}}.
\end{equation}
The last equation gives the frequency of the light emitted from a receding atom.

\subsubsection{Ives-Stilwell Experiment}

In the experiment performed by Ives and Stilwell \cite{IvesStil1,IvesStil2} (ISE),  atoms are emitted in a cathode tube and the light from the moving atoms interferes with the light from the atoms that are at rest with respect to the experimental frame (cathode tube). Clearly, the photons that interfere in ISE can be emitted both alongside the motion or against it. Let us begin with the former case.  According to the above derivations, the frequencies of the waves emitted by the speeding atoms and of the atoms at rest with respect to the cathode tube, are respectively
\begin{equation}
\omega_e=  \omega \frac{\sqrt{(1-u/v)}}{\sqrt{(1+u/c)}},
\quad \omega_r= \omega \frac{\sqrt{(1-v/c)}}{\sqrt{(1+v/c)}}.
\label{eq:catode}
\end{equation}
where $v$ is the \emph{unknown} component of Earth's relative speed that is parallel with the axis of the cathode tube; $q$ is the \emph{known} relative speed of the atoms. Then within the cathode tube,  the total speed of the moving atoms is $u=v+q$. Let us introduce a small parameter $\varepsilon = \max{(|u/c|, |v/c|, |q/c|)}$. Neglecting the terms of order $O(\varepsilon^3)$ one gets
\begin{multline}
\frac{\omega_e}{\omega_r} = \frac{\sqrt{(1-u/v)(1+v/c)}}{\sqrt{(1-v/c)(1+u/c)}} =
\left(1 -\frac{u^2}{c^2}\right)^{\frac{1}{2}}\left(1+\frac{u}{c}\right)^{-1} \left(1 -\frac{v^2}{c^2}\right)^{-\frac{1}{2}}\left(1+\frac{v}{c} \right)
\nonumber \\
= \left(1 - \frac{u}{c} + \frac{u^2}{2c^2}+ O\big(\varepsilon^3\big)\right) 
\left(1 +\frac{v}{c} + \frac{v^2}{2c^2}+ O\big(\varepsilon^3\big)\right) 
\nonumber \\
= 1 - \frac{u-v}{c} + \frac{u^2 +v^2 -2 uv}{2c^2} +  O\big(\varepsilon^3\big)
=1 - \frac{q}{c} + \frac{1}{2}\frac{q^2}{c^2} +  O\big(\varepsilon^3\big).
\end{multline}
Within the same order of approximation $O(\varepsilon^2)$, one has
\begin{subequations}\label{eq:ise}
\begin{equation}
\frac{\Delta \omega}{\omega_r}\equiv\frac{\omega_e-\omega_r}{\omega_e} = -\frac{q}{c} + \frac{1}{2}\frac{q^2}{c^2}
\ \ \Rightarrow \ \ \frac{\Delta \lambda}{\lambda} =  \frac{q}{c} + \frac{1}{2}\frac{q^2}{c^2}.
\label{eq:beat}
\end{equation}
Respectively,  for the light emitted backwards from the atoms the interference gives
\begin{equation}
\frac{\Delta \omega}{\omega_r} = \frac{q}{c} + \frac{1}{2}\frac{q^2}{c^2},\qquad
\frac{\Delta \lambda}{\lambda} =  -\frac{q}{c} + \frac{1}{2}\frac{q^2}{c^2}.
\end{equation}
\end{subequations}

The last formulas, Eqs.~\eqref{eq:ise},  show that the relative motion of Earth, as parameterized by $v$ does not affect the final result. There is a first order Doppler effect connected with the relative speed $q$ of the atoms in the cathode tube, and it is of different sign for the forward and backward emitted waves. Such `splitting' of the spectral lines was observed in \cite{IvesStil1}.  After the subtraction of the first order effect, the second order effect for both the backward and forward waves is represented by the term $\frac{1}{2}q^2$,  which is quantitatively the same as the one predicted by the hypothesis of time dilation.  It is important to understand that the first-order term detected in the ISE already speaks in favor of  a material substrate. However, the absence of dependence on the Earth's speed in the ISE was considered as another proof that the absolute medium does not exist and that the only explanation of the terms $\frac{1}{2}q^2$ can be the time dilation as predicted from the LT. We have just shown that the properly understood absolute continuum give us exactly the same frequency change as the supposed time dilation. 

The absolute motion is very elusive, indeed. It cannot be detected, in principle, by an experiment of the above type (ISE), nor by MME.  It comes as no surprise that it has not been detected in a laboratory setting. The result from this subsection claims that if properly understood, the presence of a luminiferous medium can explain the results of the Ives-Stilwell experiment, and it should be given proper consideration alongside with the hypothesis of time dilation. This conclusion lines up with the fact that electrodynamics can be explained as the manifestation of the internal stresses in an absolute continuum (called \emph{metacontinuum} in \cite{Chri_FP,Chri_MATCOM09}).

\subsection{Reconsidering the Interferometry}

Clearly, the experiment dealing with the relative speed has to be based on some superposition of beams that travel differently through the resting medium (or the LSR). Yet, interferometry based on phase difference \cite{Tolansky} does not fit the bill. Since Michelson's time split beam phase interferometry has been considered as  the only feasible approach to the problem, because it was not possible to create reasonably `identical' sources of light at two different points.  But nowadays, with the advent of highly stabilized lasers, creating almost identical sources in different spatial positions, has become a real possibility. Having two beams that travel in opposite directions, gives the opportunity to measure their beat, which allows us to call such kind of approach the `Beat-Wave Interferometry" or BWI.

The idea to use two sources of light in interferometry to create a beat frequency was floated first in \cite{Chri_CMDS8}. After the issues of the reflection from the moving mirrors were clarified in \cite{Chri_PP}, the scheme of the experiment was modified in order to avoid the cancelation of the Doppler effect. Here, we present a refined version of the theory of the new type of experiment, and clarify the issues connected with understanding the relative motion of the equipment with respect to the absolute medium.

\subsubsection{Beat Frequency from Two Opposite Beams  from Co-moving Sources}  

Assume now that two waves of almost identical frequencies are excited at two \emph{different} points that are  moving together in the same direction (say, from left to right) with the same velocity relative to the supposed medium at rest. The wave from the left source propagates in the direction of the motion (right), while the wave emitted from the right source propagates in the opposite direction (left).
 
Assume, for definiteness, that the left light source has a larger amplitude. This can happen either when it is more powerful than the other, or when there is a difference in polarizations of the sources, which is responsible for unequal amplitudes.  Denote by $f$ the respective component of either the electric or the magnetic vector.  Then the interference between the right-going wave from the left source and the left-going wave from the right source (accounting also for the Doppler effect) is given by
\begin{multline}
f(x,t) =
A_l\cos[\omega (\frac{x}{c}-t)/(1-\frac{u}{c})] +A_r \cos[\omega(1+\delta) (t + \frac{x}{c})/(1+\frac{u}{c}) +\theta]\nonumber\\
=(A_l-A_r)\cos[\omega (\frac{x}{c}-t)/(1-\frac{u}{c})]  + A_r f_1(x,t), 
\end{multline}
where $\delta$ gives the relative difference in the frequency of the second (right) source in comparison with the frequency of the first (left) source. Respectively $\theta$ gives the phase of the signal emitted from the second laser relative to the first one.  The assumption is that the polarization of the light from the two sources is not close to orthogonal, when one of the coefficients $A_l,A_r$ will be much smaller than the other. For orthogonal polarization, no interference and beat would occur. For definiteness, we choose $A_l > A_r$. In the above formula we have represented the compound wave as a undisturbed signal with amplitude $A_l-A_r$ and an interfered wave of amplitude $A_r$

For the interfered wave function $f_1$, we get
\begin{multline}
f_1(x,t) = \cos[\omega (\frac{x}{c}-t)/(1-\frac{u}{c})] 
+ \cos[\omega(1+\delta) (t + \frac{x}{c})/(1+\frac{u}{c})+ \theta] \\ =
 2 \cos[(-\hat \omega t + \tilde \omega \frac{x}{c})
 +\frac{1}{2}\omega \delta (t + \frac{x}{c})/(1+\frac{u}{c}) + \frac{1}{2}\theta
 ] \\
\qquad\times\cos[(- \tilde\omega t + \hat\omega \frac{x}{c})
-\frac{1}{2}\omega \delta (t + \frac{x}{c})/(1+\frac{u}{c}) - \frac{1}{2}\theta], 
\label{eq:interfere_absolute}
\end{multline}
where
\begin{equation}
\tilde \omega = \omega \left(1-\frac{u^2}{c^2}\right)^{-1}, \quad \hat \omega = \frac{u}{c}\tilde \omega.
\label{eq:beats}
\end{equation}

The last term on the r.h.s. of  eq.~(\ref{eq:interfere_absolute}) gives the carrier wave with frequency slightly deviating from the main frequency, while the first term on the r.h.s of  eq.~(\ref{eq:interfere_absolute}) is the modulation wave, whose frequency is a fraction of the main frequency proportional to  $u/c$.  This means that the modulation frequency is related to the first-order Doppler effect in the medium, and its measurement can give a quantitative estimate for the speed of the relative motion. This makes BWI radically different from Michelson Interferometry.

\subsubsection{Robustness of the Effect to Small Impurities of the Frequencies and Phases of the Light Sources}

It is accepted nowadays that the speed of the so-called Local Standard of Rest (LSR), to which the solar system belongs, is of the order of several hundred kilometers per second relative to the center of the local cluster of galaxies
\cite{CoreyWilk,SmootGoreMiller}. We can safely assume that $v\approx 300$ km/s which gives $\varepsilon\equiv u/c \approx 10^{-3}$. Unfortunately, there is no way of knowing what is the relative speed of the center of local cluster, so the speed of the LSR can only be considered as a reasonable guess about the speed with respect to the absolute medium.  For completeness, we mention that the lowest value for $\varepsilon$ is $10^{-4}$ based on the orbital speed of Earth, if there is no motion of the solar system with respect to the absolute medium. Thus the range for $\varepsilon$ to be targeted in an  experiment as the one proposed here, is $10^{-3} \gtrapprox  \varepsilon \gtrapprox 10^{-4}$.

Both carrier and beat frequencies in Eq.~(\ref{eq:interfere_absolute}) are modified by a term proportional to $\omega\delta$, which gives us the effect of the de-synchronization of the two laser's frequencies. Since we aim at an effect of first order with respect to the small parameter $\varepsilon$, it is sufficient to have $\delta=\varepsilon^2$ in order to reliably evaluate the beat frequency. i.e.,  $10^{-6} \gtrapprox  \delta \gtrapprox 10^{-8}$.

It should be noted here that the inevitable small drift in time of the laser frequencies, should not affect the result, because the beat wave number is much smaller than the wave numbers associated with the mentioned effects.  The sought stability of the frequency is well within the limits for  currently available low-power lasers.\footnote{Commercially available actively stabilized lasers have frequency bands from 500 kHz to 10 MHz. As a fraction of 500THz, this is a stabilization of order of $10^{-8}$.} 

If we limit ourselves to the above selected range of $\delta \sim O(u^2/c^2)$, we can neglect the terms proportional to $\delta$ and simplify Eq.~(\ref{eq:beats}) to the following
\begin{eqnarray}
f_1(x,t)= 
 2 \cos(-\hat \omega t + \tilde \omega \frac{x}{c}) \cos(- \tilde\omega t + \hat\omega \frac{x}{c}). 
\label{eq:interfere_absolute_1st}
\end{eqnarray}
Note that we have also neglected here the phase difference $\theta$ between the two lasers, because the latter simply acts to displace the carrier and beat waves in space (time) without having any effect on the quantitative value of the beat frequency (wave number). This is a crucial advantage of the BWI over the classical Michelson `phase' interferometry.

Before proceeding further, we should emphasize the fact that the above-described modulated wave is excited \emph{in the} surrounding medium which is at  \emph{rest}, rather than in the moving frames. In most of the theoretical works on the subject, this fact is very often left without comments. It is important to understand that light is emitted by elements of the moving frame, but after that it `detaches' from the latter and propagates in the absolute medium. Then it is captured again by sensors in the moving frame. The capturing process is subject to the receiver's Doppler effect. Since the entire set-up is moving with speed $u$ in the positive direction along the $x$-axis, then we have to change to a moving frame in order to get a `reading' of the pattern that is created in the absolute medium; in other words, we will have a receiver's Doppler effect at the detecting screen.  To elucidate this point, we consider the moving frame $x=\xi + ut$ and render  Eq.~\eqref{eq:interfere_absolute_1st} to the following 
\begin{equation}
f_1(\xi,t) = 2 \cos[\tilde\omega (1-\frac{u^2}{c^2})\> t -\hat\omega \frac{\xi}{c}]
\cos{(\tilde\omega \frac{\xi}{c})},
\label{eq:interfere_moving}
\end{equation}which is a standing wave $\cos{(\tilde\omega \frac{\xi}{c})}$ in the moving frame, over which propagates an envelope $\cos[\tilde\omega (1-\frac{u^2}{c^2})\> t -\hat\omega \frac{\xi}{c}]$. The essential result here is that the spatial wave number of the envelope depends on the relative velocity. This gives a unique opportunity to measure the latter. 

The difficulty stems from the fact that the phase speed of the envelope is very large, and an instantaneous snapshot can be informative only if the wave did not move appreciably in spatial direction and did not smear the beat pattern. A  feasible approach to the detection is to measure a two-point correlation. If one places several photodetectors in the area of the interaction of the two beams, one can get the correlation by time averaging the product of the two amplitudes measured. Actually there will be also a spatial averaging because of the size of the photodetector. This size has to be large enough in  comparison with the wave length of the carrier frequency, and small enough in comparison with the wave length of the beat wave. For visible light this places the dimension of the photodetectors at  $d \approx 10\mu$m. The same number gives a good estimate of the distance between the different photodetectors.

The above argument can be formalized if one introduces the following average procedure:
\begin{equation}
\langle \Phi(x,t) \rangle = \int_0^T \int _0^d \Phi \textrm{d} x\textrm{d}t.
\end{equation}
This kind of averaging will filter the highly oscillatory patterns related to the carrier wave and to the propagation speed of the envelop. Then for the correlation of the signal between two photodetectors separated by a distance $z$, we get
\begin{equation}
K(z) = \langle f(\xi+z,t)f(\xi,t) \rangle =  B \cos (\frac{uz}{c}), \label{eq:correl}
\end{equation}
where $B$ is a constant. From the profile of the correlation, one can identify the period $L$ of the cosine function $\cos(2\pi z/L)$ that fits it best. From $L$,  the relative velocity of the moving frame is identified as 
\begin{equation}
u = \frac{2\pi c^2}{2L\omega} + O(\varepsilon^2) \approx \frac{c\lambda}{2L}.
\end{equation}
For instance, if we use red-light lasers with wavelength $\lambda \approx 600 $nm, and we measure a spatial period  $L\approx 0.3$mm, then for the relative speed we get  $u \approx 300$km/s. The same result will be reached if one observes strips of width $0.6$mm when using a Infra-Red laser (IR) with wave length $1200$nm.

\subsubsection{A Possible Experimental Set-Up for First-Order Doppler effect}

The beat frequency is expected to be about $10^{-3}$ times smaller than the carrier frequency, putting it in the order of several hundred GHz.  Apart from the fact that mirrors were used in \cite{JasedaJavanTownes,JasedaJavanMurrayTownes} (see the discussion about the reflection from moving mirrors in \cite{Chri_PP}), the high beat frequency could have been another reason why it was not detected as an unwanted disturbance in those experiments. In fact, Townes et al. \cite{JasedaJavanTownes,JasedaJavanMurrayTownes} were after the much lower beat frequency connected with the second-order effects and found practically no beat. The same experiment was further refined in \cite{BrilletHall}, increasing the sensitivity 4000 times, and no beat was found. This is exactly what is to be expected in the light of the discussion from \cite{Chri_PP} where the claim was made that no effect (neither first-, second- nor higher-order) for the beat frequency can exist if reflections from \emph{moving} mirrors are involved.

The interference of light from two lasers was used in \cite{Muller} to test the isotropy of speed of light. An exclusive level of accuracy was reached there, and shown that the beat frequency is of order of a dozen of Hertz which confirms the isotropy of speed of light with accuracy better than $10^{-15}$. It also confirms the above mentioned theoretical result from \cite{Chri_PP} which states that if the source and the mirror are moving \emph{together} with respect to the absolute continuum, the Doppler effect is lost after the reflection of light from the mirror. If there is no Doppler effect, there is no beat frequency.

\begin{figure*}[h!]
\begin{center}
{\includegraphics[width=0.8\textwidth]{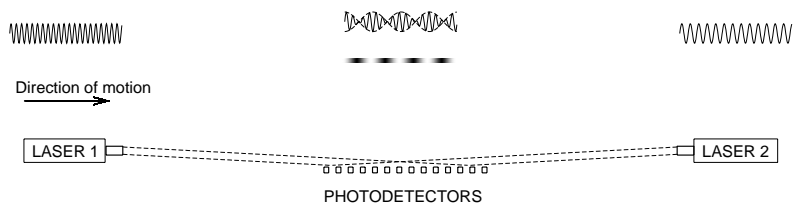}}
\caption{Experimental set-up involving two lasers/masers.}
\label{fig:two_lasers}
\end{center}
\end{figure*}

The way to conduct the experiment free of the above described difficulties connected with measuring the time frequency at a given spatial point  is to measure the wave number $\hat k$ of the spatial beat wave, by computing the above described two-point correlation function. A possible experimental set-up implementing this idea is presented in Fig.~\ref{fig:two_lasers}. As has already been pointed out, the essential feature of this experimental scheme is that it makes use of two \emph{independent} sources of coherent light. In the proposed scheme, one is supposed to observe a spatial distribution of the amplitude of the correlation as given by Eq.~\eqref{eq:correl}.
Note that using two lasers, does not make our experiment similar to the set-up used in  \cite{JasedaJavanTownes,JasedaJavanMurrayTownes}, the latter being essentially a Michelson-type interferometer designed to measure the possible change of phase. 

The available photodetectors cannot create an electric signal with a frequency of the order of the carrier frequency. The spot called `photodetectors' in Fig.~\ref{fig:two_lasers} can actually be the openings of optical fibers. The waves from two of these fibers can be mixed to produce an overall wave whose amplitude will depend on the relative phase of the waves in the two fibers. This relative phase is directly related to the beat frequency as shown in Eq.~\eqref{eq:interfere_moving}.

\section{Conclusion}

In this paper we have argued the case for existence of an absolute material continuum in which the electromagnetic vibrations propagate. The logical fallacies of the arguments that led to the so-called relativity principle are examined and the conclusion is reached that space is actually a material viscoelastic continuum which we  call the \emph{metacontinuum}. 

The linearized governing equations of the metacontinuum are shown to yield the well known Maxwell's equations as corollaries. Through judicious distinction between the referential and current descriptions, the principle of material invariance is established and shown to be a true covariance principle, unlike the Lorentz covariance, which is valid only for non-deforming frames in rectilinear relative motion. Then the new nonlinear formulation of the electrodynamics is shown to incorporate the Lorentz force, Biot--Savart, and Oersted--Ampere laws as integral parts of the model, rather than as additional empirical relations. An immediate corollary of the material invariance is the Galilean invariance of the model.

The charges (and particles in general) are considered as localized phase patterns that propagate \emph{over} the metacontinuum in the same fashion as a wave propagates over the sea surface. The motion of a material point of the metacontinuum does not have to be in the direction of the propagation of the wave; and hence no `aether wind' is to be expected. The most important property of the propagating patterns (well known from the theory of solitons) is that they shorten in the direction of motion. We show here that the contraction is proportional to the Lorentz factor, i.e. the Lorentz contraction is explained as a manifestation of the absolute continuum.

The problem of detecting the absolute continuum is also addressed. First, the Bohr-Rydberg formula for the frequency of the emitted photon is reformulated to include the effects connected with the motion of the emitting atom relative to the absolute continuum. The new formula is applied to the famous experiment of Ives and Stilwell and shown that it can be, indeed, explained from the point of view of the absolute continuum without invoking the hypothesis of time dilation. Second, a new interferometry experiment is proposed in which the first-order Doppler effect can be measured directly, and thus the Earth's relative motion with respect to the absolute continuum can be detected.

Thus a consistent model of space as an elastic material continuum has been formulated and shown to explain the observations from two of the crucial experiments. In a sense, the proposed approach can be called the `Special Theory of Absolutivity.'

\bigskip
{\bf Ackneowledgment}
The author is indebted to Dr. P. M. Jordan for numerous helpful suggestions, and for unwavering support during the last five years.


\begin{thebibliography}{11}

\parskip 1pt
\bibitem{Bland}
Bland, D. R.,
\newblock {\em Theory of Linear Viscoelasticity}.
\newblock Pergamon, New York, 1960.

\bibitem{BrilletHall}
Brillet, A. and J.~Hall.
\newblock Improved laser test of the isotropy of space.
\newblock {\em Phys. Rev. Lett}, 42:549--552, 1979.

\bibitem{Boussinesq}
Boussinesq, J. B.,
\newblock Th\'eorie nouvelle des ondes loumineuses, 
\newblock {\em J. de Math\'ematiques Pures et Applique\'es}, eries II, {\bf 13} (1868), 313--339

\bibitem{Brillouin}
Brillouin, L.,
\newblock {\em Relativity Reexamined}.
\newblock Academic Press, New York, 1970.

\bibitem{BoriTarap}
Borisenko A. I. and I. E. Tarapov,
\newblock Vector and Tensor Analysis with Applications, 
\newblock Dover, New York, 1979.

\bibitem{Chadwick}
Chadwick, P.,  
\newblock {\em Continuum Mechanics. Concise Theory and Problems}.
\newblock Dover, Mineola, New York, 1999.

\bibitem{Cantrell}
Cantrell, W.~H., 
\newblock Commentary on MaxwellÕs equations and special relativity theory,
\newblock {\em Infin. Energy}, {\bf 7}(38), (2001), 12--18.

\bibitem{Christodoulides}
Christodoulides, C.
\newblock Equivalence of the Ampere and Biot--Savart force laws in
magnetostatics, 
\newblock {\em J. Phys. A: Math. Gen.}, {\bf  20} (1987), 2037--2042.

\bibitem{Chri_WS}
 Christov, C.I., 
 \newblock On the mechanics of localized structures in continuous media, In: {\it Fluid Physics:  Proceedings of Summer Schools}, M.G.  Velarde
and  C.I.  Christov, eds., World Scientific, Singapore,
1995, pp. 33--60.


\bibitem{Chri_CMDS8}
Christov, C. I., 
\newblock Discrete out of continuous: Dynamics of phase patterns in continua.
\newblock {\em In: Continuum Models and Discrete Systems --
Proceedings of CMDS8},  K.~Markov, editor, pages 370--394, Singapore, World Scientific, 1996.

\bibitem{Chri_annuary} 
Christov, C. I.,
\newblock On the analogy between the Maxwell electromagnetic field and the elastic continuum,
\emph{Annual of University of Sofia}, \textbf{95}, (2001) 109--121.

\bibitem{Chri_ISIS_1}
Christov, C. I.,
\newblock Dynamics of patterns on elastic hypersurfaces. {Part I}. Shear Waves in the Middle Surface.
\newblock In {\em ``ISIS International Symposium on Interdisciplinary
Science'', Natchitoches, October 6-8, 2004}, pages 46--52, Washington D.C.,  APS Conference Proceedings 755, 2005.


\bibitem{Chri_ISIS_2}
Christov, C. I., 
\newblock Dynamics of patterns on elastic hypersurfaces. {Part II}. {Wave} mechanics of flexural quasi-particles.
\newblock In {\em ``ISIS International Symposium on Interdisciplinary
Science'', Natchitoches, October 6-8, 2004}, pages 53--60, Washington D.C., APS Conference Proceedings 755, 2005.

\bibitem{Chri_FP}
Christov, C.I., 
\newblock On the material invariant formulation of {Maxwell}'s displacement current.
\newblock {\em Found. Physics}, {\bf 36}, (2006) 1701--1717.

\bibitem{Chri_PP}
Christov, C.I.,
\newblock Much ado about nil: Reflection from moving mirrors and the
interferometry experiments. \newblock {\em Progress in Physics}, {\bf 3}, (2006) 55--59.

\bibitem{Chri_MATCOM07}
Christov, C. I., 
\newblock Maxwell-{L}orentz electrodynamics as a manifestation of the dynamics  of a viscoelastic metacontinuum.
\newblock {\em Math. Comput. Simul.}, {\bf 74}, (2007) 93--103.

\bibitem{Chri_WM08} Christov, C. I.,  
On the evolution of localized wave packets governed by a dissipative wave equation,  {\em Wave Motion},  {\bf 45}, (2008) 154--161. 

\bibitem{Chri_MATCOM09}
Christov, C. I.,
\newblock The concept of a quasi-particle and the non-probabilistic
  interpretation of wave mechanics.
\newblock {\em Math. Comp. Simul.}, (2009), 
\newblock To appear.

\bibitem{ChriChri_PLA}
Christov, I. and Christov, C. I.,
\newblock Physical dynamics of quasi-particles in nonlinear wave equations. \newblock {\em Phys. Lett. A}, {\bf 372},  (2008) 841--848.

\bibitem{ConsCost}
Consoli M.  and E.~Costanzo, 
\newblock From classical to modern ether-drift experiments: the narrow window
  for preferred frame.
\newblock {\em Physics Letters A}, 333:355--363, 2004.

\bibitem{CoreyWilk}
Corey B. E. and D.~T. Wilkinson, 
\newblock A measurment of the cosmic microwave background anysotropy at
  19{GHz}.
\newblock {\em Bull. Astron. Astrophys. Soc.}, 8:351, 1976.

\bibitem{Dmitriev}
Dmitriev, V. P.,
\newblock Electrodynamics and elasticity,
\newblock {\em Am. J. Phys.}, {\bf 71} (2003), 952--953.

\bibitem{Ein1905}
Einstein A.,
\newblock Zur elektrodynamik bewegeter k\"orper.
\newblock {\em Ann. der Phys.}, 17:891--921, 1905.

\bibitem{Einstein2}
Einstein A.,
\newblock {\em Relativity. The Special and the General Theory}.
\newblock Three Rivers Press, New York, 1961.

\bibitem{ErnstHsu}
Ernst, A. and J. P. Hsu, 
\newblock First proposal of the universal speed of light by {Voigt} in 1887. 
\newblock {\em Chinese J. Phys.}, {\bf 39}, (2001) 211--230.

\bibitem{GreenNaghdi}
Green, A. E. and P. M. Naghdi, 
\newblock 
A note on dipolar inertia. {\em Q. Appl. Math.}, {\bf 28}, (1970) 458--460.

\bibitem{Gill}
Gill, T. P., \newblock {\em The Doppler Effect}, Logos Press, London, 1965.


\bibitem{Graneau}
Graneau, P.,
\newblock Ampere and Lorentz forces,
\newblock {\em Phys. Lett. A}, {\bf 107} (1985), 235--237.

\bibitem{Griffiths}
Griffiths, D.~J., 
\newblock {\em Introduction to Electrodynamics}.
\newblock Prentice Hall, Englewood Cliffs, N.J, 1981.
\newblock 2nd Edition.


\bibitem{Harmuth}
Harmuth, H.
\newblock Correction of Maxwell's equations for signals I,
\newblock {\em IEEE Trans. Elmag. Compatibility}, {\bf 28} (1986), 250--258.


\bibitem{Hertz}
Hertz, H.,
\newblock {\em Electric Waves}.
\newblock MacMillan, London, 1900.

\bibitem{HsuHsu}
Hsu, J. P. and Hsu L.,
\newblock {\em A Broader View of Relativity. General Implications of Lorentz and Poincare Invariance},
\newblock Wold Scientific, New Jersey, 2006. Second Ed.

\bibitem{HutererTurner}
Huterer, D. and Turner, M. S., 
\newblock Prospects for probing the dark energy via supernova distance
  measurements.
\newblock {\em Phys. Rev. D}, (1999) {\bf 66}, \newblock Art. 081301.

\bibitem{IvesStil1}
Ives, H. E. and Stilwell, G. R. 1938
\newblock An experimental study of the rate of moving atomic clock.
\newblock {\em J. Optical Soc. Am.}, {\bf 28}, 215--226.

\bibitem{IvesStil2}
Ives, H. E. and Stilwell, G. R., 
\newblock An experimental study of the rate of moving atomic clock. II.
\newblock {\em J. Optical Soc. Am.}, {\bf 31}, (1941) 369--374.

\bibitem{JasedaJavanTownes}
Jaseda, T. S,  A.~Javan, A. and Townes, C. H. 
\newblock Frequency stability of he-Ne masers and measurements of length.
\newblock {\em Phys. Rev. Lett.}, {\bf 10}, (1963) 165--167.

\bibitem{JasedaJavanMurrayTownes}
Jaseda, T. S., Javan, A., Murray, J. and Townes, H. C., 
\newblock Test of special relativity or of the isotropy of space by use of infrared masers.
\newblock {\em Phys. Rev. A}, {\bf 113}, (1964) 1221--1225.

\bibitem{Jolly}
Jolly, D.,
\newblock Identity of the Ampere and Biot--Savart electromagnetic force laws,
\newblock {\em Phys. Lett. A}, {\bf 104} (1985), 231--234.

\bibitem{Joos}
Joos, G.,  
\newblock {\em Theoretical Physics}.
\newblock Dover, New York, 1986, Second Ed.


\bibitem{JordanPuri1}
Jordan, P. M. and  Puri, P., 
Exact solutions for the unsteady plane Couette flow of a dipolar fluid,
{\em Proc. R. Soc. Lond. A},  {\bf 458}, (2002) 1245--1272.

\bibitem{JordanPuri2}
Jordan, P. M and  Puri, A., 
\newblock Revisiting {Stokes} first problem for {Maxwell} fluids.
\newblock {\em Quart. Jl. Mech. Appl. Math.}, {\bf 58}, (2005) 213--227.

\bibitem{Joseph}
Joseph, D.~D., Narain, A,  and Riccius, O.,
\newblock Shear-wave speeds and elastic moduli for different
liquids. Part 1. Theory,
\newblock {em J. Fluid Mech.}, {\bf 171}  (1986), 28--308.

\bibitem{Karlsen}
Karlsen, B. U.,
\newblock Sketch of a matter model in an elastic universe, (1998), 
\newblock http://home.online.no/~ukarlsen/FirstPubl/WholePaper.html


\bibitem{LandauLif}
Landau, L. D. and Lifschitz, E. M., 
\newblock {\em Theory of Elasticity}.
\newblock Butterworth-Heinemann, 1986, 3rd edition.

\bibitem{Larmor}
Larmor J.
\newblock On a dynamical theory of the electric and luminiferous
medium, {\it Phil. Trans. Roy. Soc.}, {\bf 190} (897), 205-300.

\bibitem{Larmor_book}
Larmor J., 
\newblock {\it Aether and Matter}, 
\newblock Cambridge University Press, Cambridge, England, 1900

\bibitem{TDLee}
Lee, T. D.,
\newblock Broken symmetries and the physical vacuum.
\newblock {\em Nuclear Physics A}, 553:3c--14c, 1993.

\bibitem{Lodge_aberrat}
Lodge, O. 
\newblock Aberration problems. a discussion concerning the motion of ether near   the earth and concerning the connection between ether and gross matter; with   some new experiments.
\newblock {\em Phil. Trans. Roy. Soc. London A}, 184:727--804, 1893.

\bibitem{Lorentz92}
Lorentz, H.~A.,
\newblock {\em Zittingsverslagen der Akad. v. Wetenschappen te Amsterdam},
  1:74, 1892-3.

\bibitem{Lorentz95}
Lorentz, H.~A.,
\newblock Versuch einer theorie der elektrischen und optischen erschueinungen  in bewegten k\"orpern.
\newblock {\em Leiden}, pages 89--92, 1895.

\bibitem{Macrossan}
 Macrossan, M. N.,
\newblock A note on relativity before Einstein, {\it Brit. J. Phil. Sci.}, {\bf 37} (1986), 232--234.

\bibitem{Marsden}
Marsden, J.~D., Hughes, T. , 
\newblock {\em Mathematical Foundations of Elasticity}, 
\newblock Dover, New York, 1994.

\bibitem{JCM1}
Maxwell, J. C., 
\newblock A dynamical theory of the electromagnetic field.
\newblock {\em Phil. Trans. Roy. Soc. London}, {\bf 155}, (1865) 469--512.

\bibitem{JCM2}
Maxwell, J. C., 
\newblock On the dynamical theory of gases.
\newblock {\em Phil. Trans. Roy. Soc. London}, {\bf 157}, (1867) 49--88.

\bibitem{JCM3}
Maxwell, J. C.,
\newblock Ether.
\newblock In {\em Enciclopedae Britanica. Ninth Edition}, volume VIII,  1875, pages 568--572.

\bibitem{Michelson1}
 Michelson, A.~A., 
\newblock The relative motion of the earth and the luminiferous ether.
\newblock {\em Am. J. Sci.}, 22:120--129, 1881.

\bibitem{MiMor}
Michelson, A.~A. and E.~W. Morley.
\newblock On the relative motion of the earth and the luminiferous ether.
\newblock {\em Am. J. Sci.}, 34:333--345, 1887.

\bibitem{Muller}
Muller, H., S.~Hermann, C.~Braxmaier, S.~Schiller, and A.~Peters.
\newblock Moderns {Michelson}--{Morley} experiment using cryogenic optical
  resonators.
\newblock {\em Phys. Rev. Lett.}, 91:020401--1, 2003.

\bibitem{Oldroyd}
Oldroyd J. G., 
\newblock  On the formulation of rheological equations of state,
 \newblock  {\it Proc. Roy. Soc. A},
   {\bf 200} (1949), 523--541

\bibitem{Pathria}
Pathria, R. K., 
\newblock {\em The Theory of Relativity}.
\newblock Courier Dover, 2003.

\bibitem{PeeblesRatra}
Peebles, P. J. E. and  Ratra B., 
\newblock The cosmological constant and dark energy.
\newblock {\em Rev. Mod. Phys.}, {\bf 75}, (2003) 559--606.

\bibitem{Perlmutter}
Perlmutter, S., 
\newblock Measurements of omega and lambda from 42 high redshift supernovae.
\newblock {\em Astrophysical J.}, {\bf 517} (1999), 565--586.

\bibitem{Phipps}
Phipps, Jr, T.~E., {\em Heretical Verities: Mathematical Themes in Physical Description}, Classic Non-Fiction Library, Urbana, IL, (1986).

\bibitem{Pinheiro}
Pinheiro M. J.,
\newblock Do MaxwellÕs equations need revision? -- A Methodological Note, \newblock {\em arXiv:physics/0511103} (2005) .


\bibitem{QuintaStraughan}
Quintanilla, R.  and  Straughan, B.,
Bounds for some non-standard problems in porous flow and viscous Green--Naghdi fluids, {\em Proc. R. Soc. A}, {\bf 461}, (2005) 3159--3168.  

\bibitem{Riess}
Riess, A. G.,
\newblock Observational evidence from supernovae for an accelerating universe and a cosmological constant.
\newblock {\em Astronomical J.}, {\bf 116} (1998), 1009--38.

\bibitem{Sambursky}
Sambursky S., 
\newblock {\it Physics of the Stoics}, \newblock Hutchinson \& Co, London 1959. 

\bibitem{Sedov}
Sedov L.I., 
\newblock {\it A Course in Continuum Mechanics, vol. I and II},
 \newblock Walters--Nordhoff, Groningen, 1981.

\bibitem{Segel}
Segel, L. A., 
\newblock {\em Mathematics Applied to Continuum Mechanics}.
\newblock Dover, New York, 1987.

\bibitem{SharmaGanti}
Sharma P. and Ganti S., 
Gauge-field-theory solution of the elastic state of a screw dislocation in a dispersive (non-local) crystalline solid, {\em Proc. R. Soc. A},  {\bf 461}, (2005) 1081--1095.

\bibitem{SmootGoreMiller}
Smoot, G. F.,  Gorenstein, M. V. and Miller, R. A.,
\newblock Detection of anisotropy in the cosmic blackbody radiation.
\newblock {\em Phys. Rev. Lett.}, {\bf 39}, (1977) 898--901.

\bibitem{Stevenson}
Stevenson, P. M., 
\newblock Hydrodynamics of vacuum, 
\newblock {\em Int. J. Mod. Phys.}, {\bf 21} (2006), 2877--2903.

\bibitem{Tolansky}
Tolansky, S.,
\newblock {\em An Introduction to Interferometry}.
\newblock John Wiley \& Sons, New York, 1955.

\bibitem{Truesdell}
Truesdell, C.,
\newblock {\em Continuum Mechanics. Vol.I-IV}.
\newblock Gordon and Breach, New York, 1965.

\bibitem{Wang}
Wang, X. S., 
\newblock Derivation of MaxwellÕs equations based on a continuum mechanical model of vacuum and a singularity model of electric charges,
\newblock {\em Progr. Phys.}, {\bf 2} (2008), 111--120.

\bibitem{Whittaker}
Whittaker, E.~T.,
\newblock {\em A History of the Theories of Aether \& Electricity vol. 1}. \newblock Dover, New York, 1989.


\bibitem{Wise}
Wise, M. N.,
\newblock The flow analogy to electricity and magnetism, Part I: William Thomson's reformulation of action at a distance, 
\newblock {\em Archive for History of Exact Sciences}, {\bf 21} (1981),19--70.


\bibitem{PhysVac}
Zolotarev, V. ~F. and Shamshev B~.B.,
\newblock 
Structure and properties of the physical vacuum, {\it Russian Physics Journal}, {bf 18} (1985), 51-56.

\end{thebibliography}

\end{document}